\documentclass[final,5p,times,twocolumn,authoryear]{elsarticle}

\usepackage{amssymb}
\usepackage{booktabs}
\usepackage{multirow}
\usepackage{graphicx}
\usepackage{amsmath}
\usepackage{hyperref}
\usepackage{cleveref}
\usepackage{amsfonts}
\usepackage{rotating}
\usepackage{threeparttable}
\usepackage{subcaption}
\usepackage{orcidlink}
\usepackage{stfloats}
\usepackage[linesnumbered,ruled,vlined]{algorithm2e}

\crefname{figure}{Fig.}{Figs.}

\crefname{table}{Tab.}{Tabs.}

\crefname{equation}{Eq.}{Eqs.}

\crefname{section}{Sec.}{Secs.}

\usepackage{cleveref}

\crefname{algocf}{Alg.}{Algs.}
\usepackage{wrapfig}
\journal{Biomedical Signal Processing and Control}

\begin{document}
	
	\begin{frontmatter}
		
		\title{A General Framework for Generative Self-supervised Learning in Non-invasive Estimation of Physiological Parameters Using Photoplethysmography}

		\author[label1,label2,label3]{Zexing Zhang}
		\ead{2202303094@stu.ccut.edu.cn}
		\author[label1,label2,label3]{Huimin Lu\corref{cor1}
		\hspace{-1.5mm}$^{~\orcidlink{0000-0002-3786-2363}}$}
		\ead{luhuimin@ccut.edu.cn}
		
		\author[label1,label2,label3]{Songzhe Ma}
		\ead{2202103091@stu.ccut.edu.cn}
		
		\author[label1,label2,label3]{Jianzhong Peng}
		\ead{2202303023@stu.ccut.edu.cn}
		
		\author[label1,label2,label3]{Chenglin Lin}
		\ead{2202203031@stu.ccut.edu.cn}
		
		\author[label2]{Niya Li}
		\ead{liny@jlu.edu.cn}
		
		\author[label1,label2,label3]{Bingwang Dong}
		\ead{2202303105@stu.ccut.edu.cn}
		
		\affiliation[label1]{organization={School of Computer Science and Engineering},
			addressline={Changchun University of Technology}, 
			city={Jilin},
			state={Changchun},
			postcode={130102}, 
			country={China}}
		
		\affiliation[label2]{organization={Key Laboratory of Symbolic Computation and Knowledge Engineering of Ministry of Education},
			addressline={Jilin University}, 
			city={Jilin},
			state={Changchun},
			postcode={130012}, 
			country={China}}
		
		\affiliation[label3]{organization={Smart Health Joint Innovation Laboratory for the New Generation of AI}, 
			city={Jilin},
			state={Changchun},
			postcode={130102}, 
			country={China}}
		\cortext[cor1]{Corresponding author}
		
		\begin{abstract}
			Aligning physiological parameter labels with large-scale photoplethysmographic (PPG) data for deep learning is challenging and resource-intensive. While self-supervised representation learning (SSRL) can handle limited annotated data, the challenge lies in learning robust shared representations from vast unlabeled data and integrating various contextual cues to learn distinctive representations. To alleviate these challenges, a generative SSRL framework TS2TC is proposed to collaboratively utilize the temporal, spectrogram, and temporal-spectrogram mixed domains to explore and incorporate the unique features of PPG for universal and non-invasive physiological parameter estimation. Initially, a pretext task named Cross-Temporal Fusion Generative Anchor (CTFGA) is designed, modeling temporal dependencies and reconstructing independent segments at a coarse level to provide robust global feature extraction and local semantic contextual representation. The framework also includes sub-signals from PPG with diverse frequency scales and order derivatives reflecting hemodynamics to facilitate learning shared representations at varying semantic levels. Secondly, an advanced cognitive-inspired dual-process transfer (DPT) strategy is formulated, consisting of prior-dependent autonomous processes and posterior observation reasoning processes, to leverage the independent and integrated advantages of shared and specific representations. Furthermore, TS2TC introduces a novel bilinear temporal-spectrogram fusion method in the mixed domain, aligning latent representations from different domains, and establishing fine-grained contextual interactions at the feature level across multiple sources of information. Extensive experiments on physiological parameter estimation tasks showed that the joint performance of CTFGA and DPT outperforms standard generative learning significantly. TS2TC achieved an average 2.49\% improvement in RMSE over the current state-of-the-art estimation methods with only 10\% training data.
		\end{abstract}

		
	\end{frontmatter}

	\section{Introduction}
	\label{Introduction}
	\begin{figure}
		\centering
		\includegraphics[width=90mm]{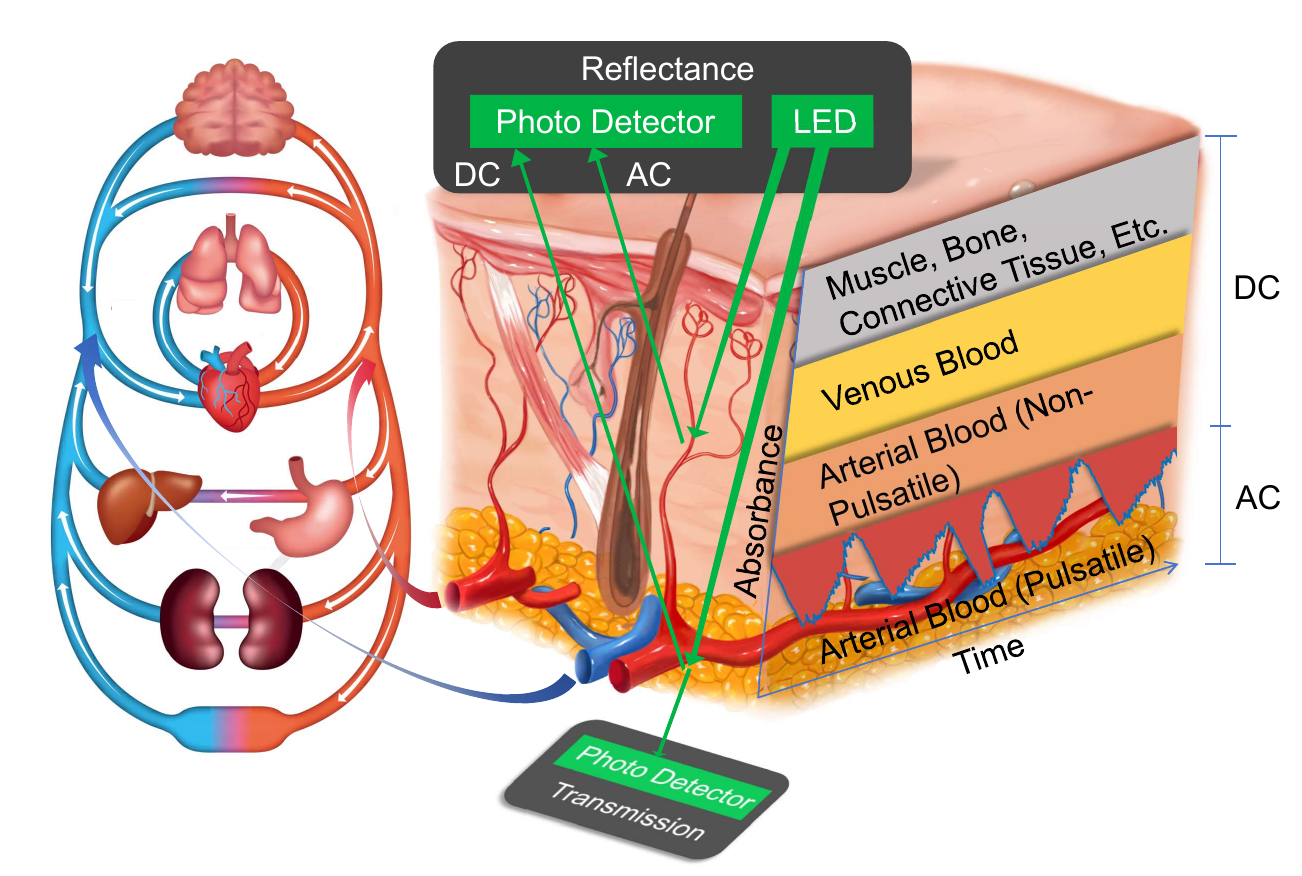}
		\caption{Overview of photoplethysmography technology.}
		\label{fig1}
	\end{figure}
	
	Photoplethysmography (PPG) is a non-invasive optical technique used to measure the absorption or reflection of light by living tissues and blood. It records pulsatile and non-pulsatile blood volumes within the covered area of the photodetector through transmission or reflectance modes, generating the raw PPG signal. The covered areas include veins, arteries, and numerous capillaries. The pulsatile component of PPG mainly reflects the complex mixture of veins and arteries in the cardiovascular system as blood flows \citep{1ray2021review}, which is highly correlated with cardiac rhythm, while the non-pulsatile component is more related to basic blood volume, sympathetic nervous system, and temperature regulation \citep{2elgendi2019use}, as shown in \cref{fig1}. Therefore, PPG holds vast potential as a new digital biomarker for non-invasive physiological parameter estimation.
	
	\begin{figure*}[hb]
		\centering
		\includegraphics[width=185mm]{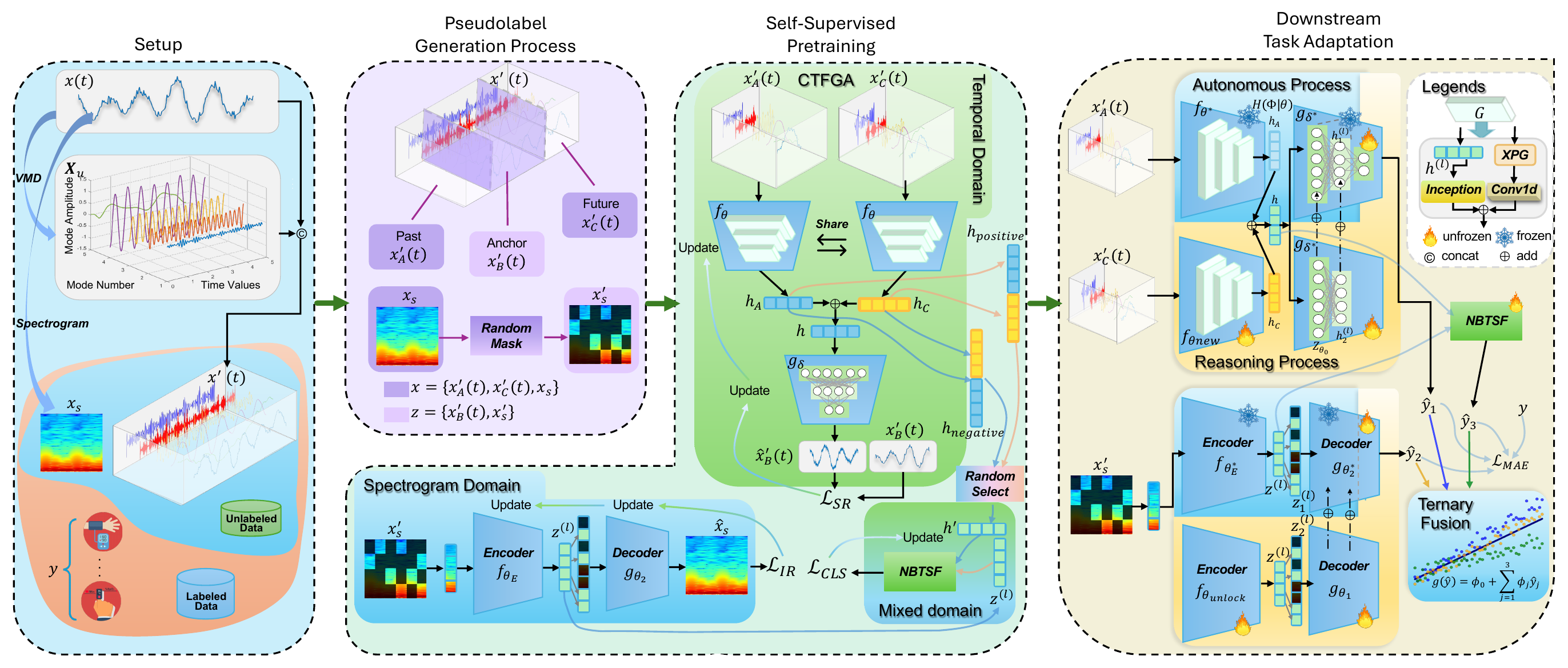}
		\caption{Overview of the TS2TC framework. In the figure, \( x(t) \) represents the input PPG signal, \( x_s \) denotes the corresponding spectrogram, and \( y \) signifies the labeled physiological parameters. \( \hat{x}_{B}'(t) \) and \( \hat{y}_{1} \) represent the signals and physiological parameters estimated in the temporal domain. Similarly, \( \hat{x}_{s} \) and \( \hat{y}_{2} \) are the spectrogram and physiological parameters estimated in the spectrogram domain, while \( \hat{y}_{3} \) represents the physiological parameters estimated in the mixed domain.}
		\label{fig2}
	\end{figure*}
	
	In recent years, the development of artificial intelligence (AI), especially deep learning (DL) algorithms, has shown promise in achieving automated, low-cost, and clinical-grade continuous estimation of physiological parameters using digital biomarkers. Convolutional neural networks (CNNs), recurrent neural networks (RNNs), transformers, and their variants are commonly used DL models for automatically extracting features and estimating physiological parameters from PPG signals. Automated feature extraction significantly surpasses traditional methods based on manually designed rules in identifying latent physiological parameters related to PPG signals. Training these models to estimate specific physiological parameters usually depends on labeled datasets aligning PPG sequences with corresponding physiological values. This supervised learning approach has excelled in applications like heart rate (HR) \citep{3kanoga2023comparison,4wang2023heart,5reiss2019deep}, blood pressure (BP) \citep{3kanoga2023comparison,6chen2022new,8kumar2023novel}, blood oxygen saturation ($\text{SpO}_{2}$)  \citep{9shuzan2023machine,10venkat2019machine,11tamura2019current}, and respiratory rate (RR) \citep{9shuzan2023machine,12charlton2017breathing,13osathitporn2023rrwavenet,14pimentel2016toward}. However, these DL models require specifically designed network architectures for each physiological parameter and do not effectively learn general and discriminative features.
	
	Existing supervised deep learning (DL) models have limitations that hinder the broader application of non-invasive physiological parameter estimation \citep{2elgendi2019use,15li2023noninvasive,16usman2017estimation,17wei2018instantaneous,18park2023association}. For example, PPG signals often lack annotated physiological parameters at the instance level, such as blood glucose (BG) which requires invasive collection \citep{15li2023noninvasive}. This scarcity leads to sparse databases, complicating the assurance of prediction accuracy, reliability, and generalizability. Fortunately, professional medical devices and ubiquitous portable devices collect large amounts of new, unlabeled PPG data daily. Thus, unsupervised learning techniques to leverage this data show promise in overcoming these limitations. Self-supervised representation learning (SSRL) is an unsupervised learning technique that extracts self-supervising information from data. Designing pretext tasks allows for the pre-training of DL models on large unlabeled databases to learn general shared representations of PPG, which can be transferred to various downstream tasks like physiological parameter regression and other specific supervised tasks. Unlike supervised DL models, SSRL requires only a small number of samples to learn specific representations while still achieving good estimation performance. SSRL currently encompasses three paradigms: contrastive learning, generative learning, and adversarial learning \citep{19zhang2024self,20krishnan2022self}. This paper introduces the TS2TC framework, which explores generative learning. 
	
	TS2 represents a combination of the temporal domain (T) using the Cross-Temporal Fusion Generative Anchor (CTFGA) method, the spectrogram domain (S) using the Masked Autoencoder (MAE) method, and the mixed domain (TS) using the novel Bilinear Temporal-Spectrogram Fusion (NBTSF) method. The '2' signifies the repeated consideration of TS and the Dual-Process Transfer (DPT) strategy. TC stands for the trinary synergy of the three domains. \cref{fig2} illustrates the TS2TC framework and its complete workflow.
	
	Specifically, in the TS2TC framework, we first reexamined the pretext task in the temporal domain of generative SSRL, aiming to learn robust and generalized shared representations during the pretraining phase. For this purpose, we introduced the concept of CTFGA to guide the self-supervised learning from signal to label (Sig2Lab). CTFGA models past and future multi-source information separately to reduce uncertainty. By reconstructing independent anchor segments, we retained the difficulty of the pretext task to learn discriminative representations while exploring temporal dependencies. Coarse-grained segment-level multi-source information helps improve the extraction ability of global features, while reconstruction helps constrain learning segment-level semantic context, creating a novel pretext task in the temporal domain. Existing methods typically rely on single past information sources for autoregressive models and lack temporal dependencies in autoencoders \citep{19zhang2024self}, as shown in \cref{fig5}. We did not follow the conventional approach of screening based on signal quality, filtering, denoising, and baseline drift removal, as these steps could lead to the loss of original semantic information. Instead, in CTFGA, we introduce noise-robust Variational Mode Decomposition (VMD) to achieve multi-frequency scale expansion of input time series. Additionally, injecting multi-order derivatives in stages guides the neural network to learn robust general features in unlabeled datasets, reducing overfitting and improving generalization. Furthermore, to integrate multi-source contextual information in downstream tasks and learn discriminative specific representations, we propose a novel domain-agnostic Dual-Process Transfer (DPT) strategy, distinct from existing fine-tuning and linear probing encoder transfer methods \citep{38zhao2023comparison}. This strategy is inspired by dual-process theories of higher cognition, where intuition and deliberate thinking correspond to two modes of human mental activity. The DPT framework divides into autonomous processes (AP) based on prior knowledge and reasoning processes (RP) based on posterior observations. A concept of a zero-decoder layer (ZDL) is introduced to facilitate interactive reasoning. Here, prior and posterior refer to shared representations learning unsupervised data distributions and specific representations learning supervised data distributions, leveraging their respective strengths through freezing and fine-tuning operations. Finally, in the context of univariate time series, the iterative bilinear temporal-spectral fusion method was expanded for the first time, exploring adaptive changes in multi-space feature fusion to enable valuable representations in mixed-domain learning. During the pre-training phase, a task for joint pre-training of temporal window relationships was designed. Subsequent tasks proposed a ternary fusion strategy based on ordinary least squares, simplifying and effectively enhancing multi-dimensional synergies and interpretability. Extensive experiments on four key physiological parameter tasks using multiple publicly available datasets demonstrated the framework's potential.
	
	To summarize, our contributions are as follows:
	
	\begin{itemize}
		\item The TS2TC framework proposes a generative SSRL framework, integrating temporal, spectrogram, and mixed domains to explore and embed intrinsic characteristics of PPG for universal and non-invasive physiological parameter estimation. To the best of our knowledge, this is the first generative self-supervised study utilizing PPG signals for estimating physiological parameters.
		\item A novel pretext task, the Cross-Temporal Fusion Generative Anchor (CTFGA), is devised where sub-signals at multiple frequency scales and dynamic derivatives from PPG signals are treated as internal features at varying semantic levels, facilitating robust shared representation learning from vast unlabeled data. This tackles the issue of balancing temporal dependencies and semantic features, along with handling erroneous contextual information from fine-grained temporal details in PPG signals.
		\item A high-level cognitive-inspired Dual-Process Transfer (DPT) strategy is formulated, incorporating autonomous processes (AP) and reasoning processes (RP), and introducing the concept of zero-decoder layers (ZDL) for reasoning interactions. By harnessing the benefits of generalized shared representations from AP and the learning advantages of RP, this approach addresses the challenge of weak specific representation capabilities when learning discriminative features with limited samples.
		\item The TS2TC framework incorporates the spectrogram domain and introduces a novel bilinear temporal-spectrogram fusion (NBTSF) method in the mixed domain, solving the alignment fusion challenge of heterogeneous and cross-domain temporal-spectrogram latent representations. Through the application of a ternary fusion strategy, the problem of unclear contributions from different sections in the complementary architecture is addressed.
	\end{itemize}
	
	The paper is structured as follows. Section 2 introduces self-supervised representation learning (SSRL) and related generative approaches. Section 3 describes the dataset and the proposed method. Section 4 presents the performance and experimental results of the TS2TC framework. Section 5 provides a detailed analysis and discussion of various components of the framework. Section 6 summarizes the research and presents concluding remarks.
	
	\section{Related work}
	\label{Related works}
	\subsection{Self-supervised representation learning (SSRL)}
	\label{SSRL}
	
	\begin{figure}
		\centering
		\includegraphics[width=70mm]{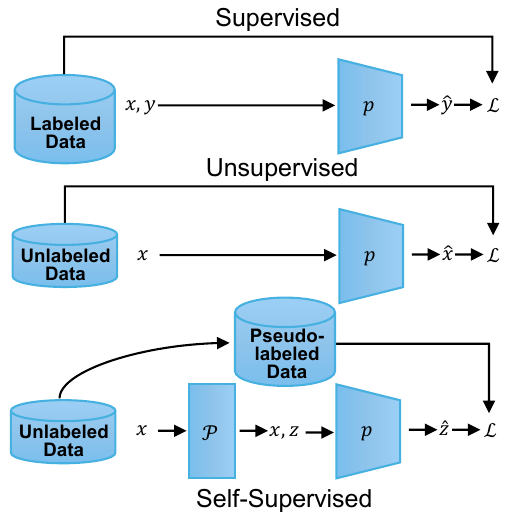}
		\caption{Contrasting supervised, unsupervised, and self-supervised learning paradigms. Unsupervised learning involves learning from unlabeled data by constructing generative models or density estimators. Here, we illustrate this with the classical method known as the autoencoder.}
		\label{fig3}
	\end{figure}
	
	SSRL, a subset of unsupervised learning, involves generating 'pseudo-labels' from data and learning valuable representations by addressing challenges encountered during the generation process (pretext task) \citep{53liu2021self,54ericsson2022self}. It has gained attention for its data efficiency and generalization capabilities, with many advanced models adopting this approach.
	
	We will compare supervised, unsupervised, and self-supervised learning paradigms to demonstrate the differences and connections between self-supervised learning and common learning paradigms, as shown in \cref{fig3}. $x$ represents the raw data, $y$ represents the labels, $\hat{y}$ is the label predicted by $p$, $\hat{x}$ is the reconstruction objective of the autoencoder, and $\mathcal{L}$ represents the loss function. By training pretext model $p$ using different learning paradigms, the acquired good representations are stored as knowledge, specifically represented by the parameters $\Theta $ of $p$. In contrast to conventional paradigms, self-supervision introduces a pretext task $\mathcal{P}$ to generate pseudo-labels $z$, used to create challenges needed for training $p$. A self-supervised pretext task is a pre-designed challenge used in self-supervised learning where the model learns from labeled data generated from the data itself without requiring human-annotated labels.
	
\begin{figure*}[bp]
		\centering
		\includegraphics[width=185mm]{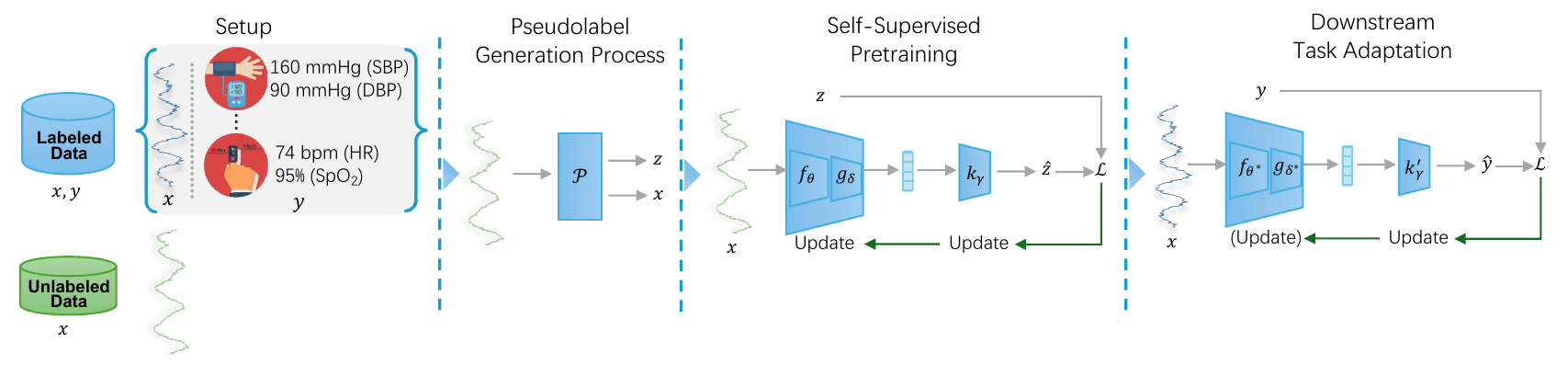}
		\caption{Workflow of self-supervised representation learning. The term '(Update)' signifies that parameter updates are optional. When not updated, the method follows a linear probing paradigm; when updated, it adheres to a fine-tuning paradigm. (SBP is systolic blood pressure, DBP is diastolic blood pressure, HR is heart rate, $\text{SpO}_{2}$ is blood oxygen saturation)}
		\label{fig4}
	\end{figure*}

	Generally, complete self-supervised representation learning includes the training process of pretext model $p$ described in \cref{fig3}, as well as the process of knowledge transfer for downstream supervised tasks. The complete workflow is shown in \cref{fig4}. Furthermore, this can be described as follows:
	
	\textbf{Setup.} Generate a limited, annotated database $D_t$ for downstream tasks such as non-invasive heart rate estimation parameters, and an accessible unlabeled database $D_s$.
	
	\textbf{Pseudolabel Generation Process.} Generate a new pseudo-labeled database $\bar{D}_{s}=\{x_i,z_i\}_{i=1}^M=\mathcal{P}(D_s)$ through the pretext task $\mathcal{P}$.
	
	\textbf{Self-Supervised Pretraining.} Training the pretext model $p=k_{\gamma}(g_{\delta}(f_{\theta}(\cdot)))$ by optimizing the self-supervised objective on $\bar{D}_{s}$. Preserve the optimized parameter set $\Theta^{*}=\{\theta^{*},\delta^{*}\}$, keeping the well-learned representational knowledge by retaining only the parameters of the encoder $f_{\delta}$ and decoder $g_{\theta}$, while disregarding the pretext output function $k_{\gamma}$:
	
	\begin{equation}
		\begin{aligned}
			\Theta^*=\underset{\theta,\delta,\gamma}{\operatorname*{minarg}}\sum_{(x_i,z_i)\in\mathcal{P}(D_s)}\mathcal{L}\big(k_\gamma(g_\delta\big(f_\theta(\cdot)\big)\big),z_i\big)
		\end{aligned}
		\label{eq1}
	\end{equation}
	
	\textbf{Downstream Task Adaptation.} Utilize the representational function $g_{\delta^{*}}(f_{\theta^{*}}(\cdot))$ and its knowledge $\Theta^{*}$ for partial transfer to address the target problem of interest, which is the downstream task. Fine-tuning and linear probing are two common paradigms for knowledge transfer and addressing target problems \citep{19zhang2024self}. The linear probing involves designing a specialized head $k_{\gamma}^{\prime}$ for a new target problem and then training this head using a frozen pretext model $p$. Here, 'frozen' implies that the representational knowledge obtained through the pretext task will not be updated. Additionally, the approach allows training a new head $k_{\gamma}^{\prime}$ and retraining the entire network for a new task, involving the simultaneous update of the model $p$ and $k_{\gamma}^{\prime}$, representing the fine-tuning method. \cref{DPT} provides further details on fine-tuning and linear probing.
	
	Existing pretext tasks can be roughly summarized into three categories: context prediction, instance discrimination, and instance generation \citep{19zhang2024self}. The strategies proposed in this paper all fall under context prediction.
	
	\subsection{Generative-based SSRL}
	\label{Generative-based SSRL}
	
\begin{figure}[h]
		\centering
		\begin{subfigure}{0.48\textwidth}
			\centering
			\includegraphics[width=\textwidth]{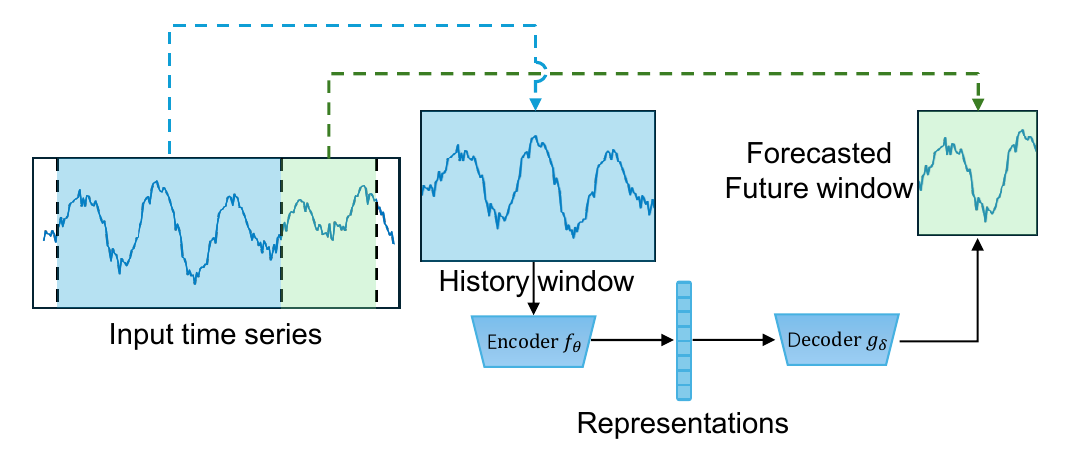}
			\caption{Autoregressive-based forecasting task}
			\label{fig5a}
		\end{subfigure}
		\hspace{1cm}
		\begin{subfigure}{0.48\textwidth}
			\centering
			\includegraphics[width=\textwidth]{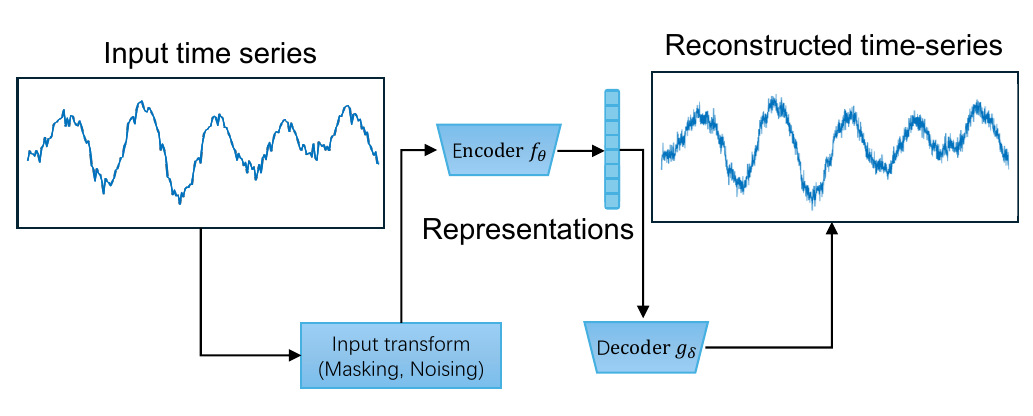}
			\caption{Autoencoder-based reconstruction task}
			\label{fig5b}
		\end{subfigure}
		\caption{Generative-based SSRL for time series data. (Repainted from \citep{19zhang2024self})}
		\label{fig5}
	\end{figure}

	According to different learning paradigms in self-supervised representation learning, we can categorize SSRL into generative-based methods, contrastive-based methods, and generative-contrastive (adversarial) methods. Here, we focus on generative-based SSRL methods. In generative-based methods, the pretext task involves generating expected data based on a given data view. In the context of time series modeling, common pretext tasks include using past sequences to forecast future windows or specific timestamps (autoregressive prediction) and using an encoder and decoder to reconstruct input (autoencoder-based reconstruction), as shown in \cref{fig5}.
	
	As our objective is to develop a general framework independent of specific physiological parameter estimation tasks, we will present the distinct features of PPG signals and related generative SSRL work, emphasizing their role as time-series attributes. In the temporal domain, where input data sources are derived from continuous time series, autoregressive (AR) models are commonly used as pretext tasks in SSRL. Unlike AR, SSTSC \citep{22xi2022semi} does not generate values for future time windows directly. Instead, it segments the time series into 'past-anchor-future' fragments, generating relationships between these windows to fully leverage the temporal relational structure in the data, thereby enhancing feature representation capabilities. This approach shifts the focus away from semantic information. STEP \citep{23shao2022pre} divides the time series into multiple equal-length, non-overlapping, fine-grained, independent fragments and then randomly selects a proportion of these for masking and reconstruction tasks. It suggests that reconstruction can capture higher-quality semantic context for short-term time series and facilitate the modeling of fine-grained dependencies. PITS \citep{24lee2023learning}, however, argues that extensively modeling fragment dependencies might not be the optimal strategy for time series representation learning. Another common pretext task in SSRL, similar to autoregression, is autoencoding \citep{18park2023association}, which independently reconstructs short-term time fragments. We balanced the reconstruction strategy by simultaneously modeling dependencies and reconstructing independent fragments.
	
	Meanwhile, PPG signals are significantly influenced by motion artifacts \citep{1ray2021review}, and fine-grained local segments may provide incorrect contextual information, especially when redundant information is present in adjacent segments \citep{52chen2023rafnet}. In the short term, physiological parameters are relatively stable. Therefore, considering the global characteristics of coarser segments can yield information gain. Variants of the transformer are widely used in time-series masking and reconstruction tasks \citep{23shao2022pre,24lee2023learning}. However, a contrastive self-supervised learning study of AHMS \citep{25abbaspourazad2023large} demonstrated that on large-scale PPG datasets, the 1-dimensional transformer variant (TSiTransformer \citep{40dosovitskiy2020image}) achieved similar encoder performance to convolutional neural networks, but the convolutional networks had significantly fewer parameters and lower computational complexity. Additionally, derived signals naturally aligned with the temporal sequence of PPG play a crucial role in the estimation of physiological parameters, particularly in estimating latent physiological parameters. For instance, decomposed sub-signals \citep{17wei2018instantaneous} and higher-order derivatives representing changes in cardiovascular circulatory dynamics \citep{16usman2017estimation,18park2023association}. BTSF \citep{21yang2022unsupervised}, in the enhanced views of the temporal and frequency domains, utilizes a bilinear temporal-frequency fusion mechanism for information integration and significantly outperforms the then state-of-the-art methods for time-series prediction. However, it is designed for multivariate time series and may lose the semantic information of transient changes and the evolution of frequency components over time in non-stationary signals. CS-NET \citep{8kumar2023novel} explored the effectiveness of the time-frequency spectrogram of PPG signals in physiological parameter prediction but overlooked the potential contribution of the original time-domain signals.
	
	This paper proposes a generative-based SSRL framework TS2TC, as shown in \cref{fig2}, which decouples from specific physiological parameter estimation tasks and is detailed in \cref{Method}.
	
	\section{Method}
	\label{Method}
	
	The transition of PPG signal analysis from the temporal domain to the frequency domain is a notable trend \citep{8kumar2023novel}. These domains are interconnected and complement each other. The conversion of dynamic signals from the temporal domain to the frequency domain is primarily achieved through the Discrete Fourier Transform (DFT). DFT decomposes mixed signals into sine and cosine waves with different frequencies to extract frequency domain information. However, DFT is suitable only for stationary signals, while PPG signals are non-stationary. Moreover, frequency domain analysis cannot capture changes in frequency over time. Thus, single-dimensional analyses in the temporal or frequency domain alone cannot comprehensively characterize PPG signals. Alongside retaining temporal domain analysis, we have introduced spectrogram domain analysis. Building on the success of temporal-frequency fusion \citep{21yang2022unsupervised}, our framework integrates the temporal domain and spectrogram domain synergistically. This integration results in the trilateral synergistic TS2TC framework, as depicted in \cref{fig2}. \cref{alg:TS2TC} outlines the complete pipeline of TS2TC, with subsequent sections providing detailed explanations of the proposed strategies.
	
	\subsection{Cross-temporal fusion generated anchor (CTFGA)}
	\label{CTFGA}
	
	The PPG signals are commonly viewed as 1-D time series with highly nonlinear and non-stationary traits. Assuming $x(t)$ represents a single-channel PPG signal, the VMD method decomposes the complex $x(t)$ into stationary sub-signals $\boldsymbol{X}_{u}=\left\{u_{i}(t)\right\}=\left\{u_{1}(t), u_{2}(t), \ldots, u_{k}(t)\right\}$ with different central frequencies $\{\omega_i\}$: 
	
	\begin{equation}
		x(t)=\sum_{i=1}^{k} u_{i}(t)+res(t)
		\label{eq1}
	\end{equation}
	
	Here, $k$ denotes the number of sub-signals, and $res$ signifies the optimized residual signal. $\boldsymbol{X}_{u}$ offers robust derived information aligned with the natural time sequence of $x(t)$, aiding in effectively mitigating the uncertainty impact from nonlinear and non-stationary elements in PPG signals. The Setup section in \cref{fig2} illustrates this process.
	
	\textbf{Variational Mode Decomposition.} VMD is an adaptive non-recursive signal decomposition technique \citep{37dragomiretskiy2013variational}, which decomposes the input signal into a series of sparse characteristic signals $\{u_i(t)\}$, constructing the decomposition as the process of solving the optimal solution of the constrained variational problem:
	
	\begin{equation}
		\left\{
		\begin{aligned}
			& \min _{u_{i}, \omega_{i}} \sum_{i}\left\|\partial_{t}\left[\left(\delta(t)+\frac{j}{\pi t}\right) * u_{i}(t)\right] e^{-j \omega_{i} t}\right\|_{2}^{2} \\
			& \text{s.t. } \sum_{i} u_{i}=x(t)
		\end{aligned}
		\right.
		\label{eq2}
	\end{equation}
	
	\begin{algorithm}[t]
		\caption{TS2TC Framework}
		\label{alg:TS2TC}
		\KwIn{PPG signal $x(t)$ with partial labels $y$, partition ratio $\gamma$, sampling rate $M$, and sampling time $T$}
		\KwOut{Predicted PPG signal labels $g(\hat{y})$}
		\BlankLine
		\textbf{Phase 1: Setup}\;
		$\textbf{\textit{X}}_{u} \gets \textit{VMD}(x(t))$ (Algorithm \ref{alg:VMD})\;
		$x^{\prime}(t) = \textit{concat}(x(t),\boldsymbol{X}_{u})$\;
		$x_s \gets \textit{spectrogram}(x(t))$\;
		\BlankLine
		\textbf{Phase 2: Pseudolabel Generation Process}\;
		$x_{A}^{\prime}(t) = x^{\prime}(t)[:, :\gamma MT]$\;
		$x_{B}^{\prime}(t) = x^{\prime}(t)[:, \gamma MT:(1-\gamma)MT]$\;
		$x_{C}^{\prime}(t) = x^{\prime}(t)[:, (1-\gamma)MT:]$\;
		$x_s^{\prime} \gets RandomMask(x_s)$\;
		\BlankLine
		\textbf{Phase 3: Self-Supervised Pretraining}\;
		$f_\theta,g_{\delta} \gets \textit{CTFGA}(x_{A}^{\prime}(t),x_{B}^{\prime}(t),x_{C}^{\prime}(t))$ (Algorithm \ref{alg:CTFGA})\;
		$\theta_E,\theta_2,z^{(l)} \gets \textit{SpectrogramDomain}(x_s,x_s^{\prime})$ (Algorithm \ref{alg:Spectrogram})\;
		
		$h_A=f_\theta(x_A^{\prime}(t))$ \quad $h_C=f_\theta(x_C^{\prime}(t))$\;
		$h_{positive}=\textit{concat}(h_{A},h_{C})$ \quad
		$h_{negative}=\textit{concat}(h_{C},h_{A})$\;
		$h^{\prime} = \textit{RandomSelect}(h_{positive},h_{negative})$\;
		
		$F_{enhance}=\textit{NBTSF}(z^{(l)}, h^{\prime})$ (Algorithm \ref{alg:NBTSF})\; $f_{\mu}\gets\mathcal{L}_{CLS}(f_{\mu}(F_{enhance_{i}}),c_{i})$ (Algorithm TWRG)\;
		\BlankLine
		\textbf{Phase 4: Downstream Task Adaptation}\;
		$\hat{y}_1 = \hat{y} \gets \textit{CTFGA w/ DPT}(f_{\theta},g_{\delta},x_{A}^{\prime}(t),y)$ (Algorithm \ref{alg:DPT})\;
		$\hat{y}_2 = \hat{y},z^{(l)} \gets \textit{SpectrogramDomain w/ DPT}(\theta_E,\theta_2,x_s^{\prime},y)$ (Algorithm \ref{alg:Spectrogram})\;
		$h^{\prime} = h_{positive} \quad F_{enhance}=\textit{NBTSF}(z^{(l)}, h^{\prime})$ (Algorithm \ref{alg:NBTSF})\;
		Remove the Softmax from $f_{\mu}$\;
		$\hat{y}_3 = f_{\mu}(F_{enhance})$\;
		Calculate losses: $\mathcal{L}_{MAE}(\hat{y}_1,y)\quad\mathcal{L}_{MAE}(\hat{y}_2,y)\quad\mathcal{L}_{MAE}(\hat{y}_3,y)$\;
		Backpropagation: Compute gradients and update unfrozen parameters\;
		$\phi \gets \textit{TernaryFusion}(\hat{y}_1,\hat{y}_2,\hat{y}_3,y)$ (\cref{eq25,eq26})\;
		\BlankLine
		\textbf{Phase 5: Inference}\;
		$g(\hat{y}) \gets \textit{TernaryFusion}(\hat{y}_1,\hat{y}_2,\hat{y}_3,\phi)$ (\cref{eq25})\;
	\end{algorithm}
	
	\begin{algorithm}[h]
		\caption{Variational mode decomposition (VMD)}
		\label{alg:VMD}
		\SetAlgoLined
		\KwIn{Signal \( x(t) \), penalty parameter \( \alpha \), noise tolerance \( \tau \), convergence error \( \varepsilon \)}
		\KwOut{Modes \( \textbf{\textit{X}}_u \)}
		\BlankLine
		\textbf{Initialize:} \( u_i^0(t), \omega_i^0, \lambda^0(t) \)\;
		\Repeat{convergence \textnormal{(\cref{eq7})}}{
			\For{each mode \( i \)}{
				Compute mode \(\hat{u}_i^{n+1}(\omega)\) in frequency domain using \cref{eq4}\;
				Update \(\omega_i^{n+1}\) using \cref{eq5}\;
				Update Lagrange multiplier \(\lambda^{n+1}(\omega)\) using \cref{eq6}\;
			}
		}
		\For{each mode \( i \)}{
			Compute \( u_i(t) \) using inverse Fourier transform\;
			\[
			\hat{u}_i(t) = \mathcal{F}^{-1}(\hat{u}_i(\omega))
			\]
		}
		
		Assign \( \{\hat{u}_i(t)\} \) to \( \textbf{\textit{X}}_{u} \)\;
		\[
		\textbf{\textit{X}}_{u} = \{\hat{u}_{1}(t),\hat{u}_{2}(t), \ldots,\hat{u}_{k}(t)\}
		\]
	\end{algorithm}
	
	$\delta(t)$ is an impulse function, Wiener filtering is integrated into the VMD algorithm for enhanced robustness against sampling and noise. \cref {eq2} can be solved by introducing a quadratic penalty $\alpha$ and Lagrange multiplier $\lambda$. The augmented Lagrange equation is given as follows:
	
	\begin{equation}
		\begin{split}
			L(\{u_{i}\},\{\omega_{i}\},\lambda)=&\alpha\sum_{i=1}^{k}\left\|\partial_{t}\left[\left(\delta(t)+\frac{j}{\pi t}\right)*u_{i}(t)\right]e^{-j\omega_{i}t}\right\|_{2}^{2}+\\
			&\hspace{-1cm} \left\| x(t)-\sum_{i=1}^{k}u_{i}(t)\right\|_{2}^{2}+\langle\lambda(t),x(t)-\sum_{i=1}^{k}u_{i}(t)\rangle
			\label{eq3}
		\end{split}
	\end{equation}

	Where $\alpha$ is the trade-off parameter for data fidelity constraint. Through iterations, $\{\omega_i\}$ and $\{\hat{u}_{i}(\omega)\}$ in the frequency domain can be estimated separately:

	\begin{equation}
		\hat{u}_{i}^{n+1}(\omega)\leftarrow\frac{\hat{x}(\omega)-\sum_{k<i}\hat{u}_{k}^{n+1}(\omega)-\sum_{k>i}\hat{u}_{k}^{n}(\omega)+\frac{\lambda^{n}(\omega)}{2}}{1+2\alpha\big(\omega-\omega_{i}^{n}\big)^{2}}
		\label{eq4}
	\end{equation}
	
	\begin{equation}
		\omega_{i}^{n+1}\leftarrow\frac{\int_{0}^{\infty}\omega\left|\hat{u}_{i}^{n+1}(\omega)\right|^{2}d\omega}{\int_{0}^{\infty}\left|\hat{u}_{i}^{n+1}(\omega)\right|^{2}d\omega}
		\label{eq5}
	\end{equation}
	
	$\omega_i$ is calculated at the centroid of the corresponding sub-signal power spectrum, and after updating $\hat{u}_{i}(\omega)$ and $\omega_i$ every time, $\lambda$ is also iteratively updated:
	
	\begin{equation}
		\lambda^{n+1}(\omega)\leftarrow\hat{\lambda}^{n}(\omega)+\tau\left(\hat{x}(\omega)-\sum_{i}\hat{u}_{i}^{n+1}(\omega)\right)
		\label{eq6}
	\end{equation}
	
	The iteration termination criterion is:
	
	\begin{equation}
		\sum_i\frac{\parallel\hat{u}_i^{n+1}-\hat{u}_i^n\parallel_2^2}{\parallel\hat{u}_i^n\parallel_2^2}<\varepsilon 
		\label{eq7}
	\end{equation}
	
	Where $\tau$ is the noise tolerance, and $\varepsilon$ is the convergence error, $\{\hat{u}_{i}(\omega)\}$ inverse Fourier transform $\mathcal{F}^{-1}$, converted to the temporal-domain modal components $\{\hat{u}_{i}(t)\}$. \cref{alg:VMD} describes the process of computing $\textbf{\textit{X}}_{u}$ in the VMD method. Finally, $x(t)$ and $\textbf{\textit{X}}_{u}$ are combined into an expanded signal of multi-frequency scales $ x^{\prime}(t)=\{x(t),\hat{u}_{1}(t),\hat{u}_{2}(t), \ldots,\hat{u}_{k}(t)\}$, the fusion representing the concat operation in \cref{fig2}.
	
	\textbf{Cross-Temporal Fusion Generative Anchor.} \cref{alg:CTFGA} describes the complete CTFGA strategy process. First, the inflated signal $x^{\prime}(t)\in\mathbb{R}^{k\times (MT)}$ is partitioned in time by the factor $\gamma$ to form $x_{A}^{\prime}(t)\in\mathbb{R}^{k\times (\gamma MT)}$ representing past information sources and $x_{C}^{\prime}(t)\in\mathbb{R}^{k\times (\gamma MT)}$ representing future information sources. Finally, the anchor information $x_{B}^{\prime}(t)\in\mathbb{R}^{k\times ((1-2\gamma)MT)}$ is predicted using $x_{A}^{\prime}(t)$ and $x_{C}^{\prime}(t)$. $M$ and $T$ are sampling rate and sampling time, respectively, with a fixed value. The Pseudolabel Generation Process section in \cref{fig2} illustrates this. Additionally, suitable $\gamma$ and $T$ values were investigated and chosen in \cref{Exploration of CTFGA's best γ and T}.
	
	Given a neural network encoder $f_{\theta}\colon x\in\mathbb{R}^{k\times (\gamma MT)}\to h\in\mathbb{R}^{E}$, which maps on $k\times (\gamma MT)$ dimensional input $x$ to an $E$-dimensional latent space representation $h$, where $\theta$ represents the encoder parameters. The function $f_{\theta}(\cdot)$ can map $x_{A}^{\prime}(t)$ and $x_{C}^{\prime}(t)$ to representations $h_{A}=f_{\theta}(x_{A}^{\prime})$ and $h_{C}=f_{\theta}(x_{C}^{\prime})$. PPG signals generate the first derivative $vpg(t)$, the second derivative $apg(t)$, and the third derivative $jpg(t)$, which, from a physical perspective, reflect the speed, acceleration, and jerk of blood, respectively:
	
	\begin{equation}
		\left\{
		\begin{aligned}
			& vpg(t) = \frac{\mathrm{d}}{\mathrm{d}t}\big(x(t)\big) = \frac{\mathrm{d}}{\mathrm{d}t}\big[x(t_{i})-x(t_{i-1})\big]\\
			& apg(t) = \frac{\mathrm{d}}{\mathrm{d}t}\big(vpg(t)\big) = \frac{\mathrm{d}}{\mathrm{d}t}\big[x(t_{i})-2x(t_{i-1})+x(t_{i-2})\big]\\
			& \begin{aligned}
				jpg(t) &= \frac{\mathrm{d}}{\mathrm{d}t}\big(apg(t)\big) \\
				&= \frac{\mathrm{d}}{\mathrm{d}t}\big[x(t_{i})-3x(t_{i-1})+3x(t_{i-2})-x(t_{i-3})\big]
			\end{aligned}
		\end{aligned}
		\right.
		\label{eq8}
	\end{equation}
	
	From a feature perspective, $vpg(t)$, $apg(t)$ and $jpg(t)$ represent high-level features at different levels. Injecting them as internal features at different stages into the network can enhance the generalization ability of $f_{\theta}(\cdot)$, allowing for the learning of features that better fit the continuous modeling process. Using PPG derivatives as internal features is considered a guide block $G$ (as shown in the legend of \cref{fig2}), and the guiding process is as follows:
	
	\begin{equation}
		\begin{aligned}
			h^{(l+1)} &= G(h^{(l)}) \\
			&= \text{BatchNorm}(\text{Conv1d}_i^{(l+1)}(xpg^{(l)})) \\
			&\oplus \text{NetBlock1d}_i^{\prime(l+1)}(h^{(l)})
		\end{aligned}
		\label{eq9}
	\end{equation}
	
	Where $Conv1d_{i}^{(l+1)}=\sum_{p=1}^{s}\chi_{i+p-1}^{(l)}\cdot k_{p}^{(l+1)}+b^{(l+1)}$ represents 1-D convolution operation when the input feature at layer $l$ is $\chi^{(l)}$, and the convolution kernel is $k^{(l+1)}$. Introducing PPG derivatives $xpg\in\{vpg,apg,jpg\}$ as injected elements, the element-wise summation operation $\oplus$ represents external information-guided continuous modeling. Different from internal feature injection like the residual network to alleviate gradient vanishing and enhance network feature extraction performance, here external information guides continuous modeling and introduces additional rich information to improve network performance. Moreover, $NetBlock1d_{i}^{\prime(l+1)}$ represents different style encoder blocks, this study adopts the Inception \citep{42tan2021time} encoder block style that has performed well in PPG-related tasks for development, considering its use of filters of different lengths and shortcut skip connections. The processing of the $NetBlock$ can be described as follows:
	
	\begin{equation}
		NetBlock1d_{i}^{\prime(l+1)}(x) = \begin{aligned}[t]
			&ReLU\left[BatchNorm(Conv1d_{i}^{\prime(l+1)}(x) \right.\\
			&\left. \oplus Bottleneck(x))\right]
		\end{aligned}
		\label{eq10}
	\end{equation}
	
	\begin{algorithm}[h]
		\caption{Cross-Temporal Fusion Generated Anchor (CTFGA)}
		\label{alg:CTFGA}
		\SetAlgoLined
		\KwIn{Dilated signal $x_{A}^{\prime}(t),x_{B}^{\prime}(t),x_{C}^{\prime}(t)$}
		\KwOut{Trained neural network encoder $f_{\theta}$ and decoder $g_{\delta}$}
		\BlankLine
		\textbf{Initialize:} Encoder parameters $\theta$ and decoder parameters $\delta$\;

		\Repeat{not converged}{
			\For{each layer $l$ in encoder}{
				$h^{(l)}_{A}, h^{(l)}_{C} \leftarrow f_{\theta}^{(l)}(x_{A}^{\prime}), f_{\theta}^{(l)}(x_{C}^{\prime})$\;
				
				\eIf{Layer $l$ is one of the first three layers}{
					\eIf{Layer $l$ is the first layer}{
						$xpg^{(l)} = vpg^{(l)}(t) = \frac{\mathrm{d}}{\mathrm{d}t} x^{(l)}(t)$\;
					}{
						\eIf{Layer $l$ is the second layer}{
							$xpg^{(l)} = apg^{(l)}(t) = \frac{\mathrm{d}}{\mathrm{d}t} vpg^{(l)}(t)$\;
						}{
							$xpg^{(l)} = jpg^{(l)}(t) = \frac{\mathrm{d}}{\mathrm{d}t} apg^{(l)}(t)$\;
						}
					}

					Update $h^{(l+1)}$ using \cref{eq9}\;
				}{
					Continue without guide block\;
				}
			}
			$h = \text{concat}(h_{A}, h_{C})$\;
			$\hat{x}_{B}^{\prime} \leftarrow g_{\delta}(h)$\;
			
			\textbf{Calculate Loss:}
			\[
			\mathcal{L}_{SR} = \|\hat{x}_{B}^{\prime}(t) - x_{B}^{\prime}(t)\|_{1}
			\]
			
			\textbf{Backpropagation:} Compute gradients and update $\theta$ and $\delta$\;
		}
		
	\end{algorithm}
	
	$Conv1_{i}^{\prime(l+1)}(x)$ represents the composite of three different scale convolution operations, and the $Bottleneck$ layer performs operations with convolution filters of length and stride, both equal to 1. $BatchNorm(Conv1d_{i}^{(l+1)}(xpg))$ and $BatchNorm(Conv1d_{i}^{\prime(l+1)}(x))$ represent batch normalization operations, and the convolution output $z^{(l+1)}$ following the operations will follow a normal distribution:
	
	\begin{equation}
		\left\{
		\begin{aligned}
			& \mu_{\beta} \leftarrow \frac{1}{m}\sum_{i=1}^{m}z^{(l+1)} \\
			& \sigma_{\beta}^2 \leftarrow \frac{1}{m}\sum_{i=1}^{m}(z^{(l+1)}-\mu_{\beta})^{2} \\
			& \hat{z}^{(l+1)} \leftarrow \frac{z^{(l+1)}-\mu_{\beta}}{\sqrt{\sigma_{\beta}^{2}+\varepsilon}}
		\end{aligned}
		\right.
		\label{eq11}
	\end{equation}
	
	Next, the latent space representation information is fused into $h=\{h_{A},h_{C}\}$, then consider the neural network decoder $g_{\delta}\colon x\in\mathbb{R}^{2E}\to h\in\mathbb{R}^{k\times ((1-2\gamma)MT)}$, where $\delta$ represents the parameters of the decoder, decoding the latent space representation $h$ into the anchor information $x_{B}^{\prime}(t)$. This maps cross-temporal continuous information to the latent space, covering past and future relevant features, including advanced features of independent continuous information, rather than local discrete features. The decoder $g_{\delta}$ is uniformly composed of a multi-layer perceptron, including input and output layers as well as three hidden layers. In addition, to reduce overfitting, there is a dropout layer between two adjacent hidden layers, with a dropout rate set to 0.1 for both. Assuming variables of layer $l$ are $h^{l}$, the output of layer $l+1$ can be represented as:

	\begin{equation}
		\left\{
		\begin{aligned}
			& a^{(l+1)} = W^{(l)}h^{(l)} + b^{(l)} \\
			& h^{(l+1)} = \psi^{(l+1)}\bigl(a^{(l+1)}\bigr)
		\end{aligned}
		\right.
		\label{eq12}
	\end{equation}
	
	$b^{l}\in\delta$ and $W^{l}\in\delta $ are learnable bias terms and weight parameters respectively, and $\psi(\cdot)$ represents the dropout layer operation. Parameters $\theta$ and $\delta$ are updated by minimizing the $\ell_{1}-norms$ loss $\mathcal{L}=\|\hat{x}_{B}^{\prime}(t)-x_{B}^{\prime}(t)\|_{1}$, where $\hat{x}_{B}^{\prime}(t)$ denotes the predicted anchor information. The signal $x_{B}^{\prime}(t)$ serves as the ground truth anchor signal obtained from the partitioning of $x^{\prime}(t)$. The overall process of CTFGA is detailed in the Self-Supervised Pretraining section of \cref{fig2}.
	
	\subsection{Dual-process transfer (DPT)}
	\label{DPT}
	
	Fine-tuning and linear probing are two paradigms for the transfer of SSRL encoders \citep{38zhao2023comparison}. In the fine-tuning strategy, model training follows a two-phase approach. Initially, the encoder $f_{\theta}$ learns valuable latent representations $h$ from a large unlabeled PPG dataset through pretext tasks, with $\theta$ serving as the initial parameters for the new encoder $f_{\theta new}(\cdot)$. In the second phase, further training involves fine-tuning the pre-trained encoder $f_{\theta new}(\cdot)$ parameters on a relatively smaller labeled dataset. This process relies on the new linear prediction head $g_{new}(\cdot)$:
	
	\begin{equation}
		y'=g_{new}(h)=Wh+b=Wf_{\theta new}\big(x(t)\big)+b
		\label{eq13}
	\end{equation}
	
	Here, $W$ and $b$ respectively represent the weights and biases of the linear regression head. Conversely, linear probing keeps $\theta$ frozen in the second phase and updates only the parameters of $g_{new}(\cdot)$ through training.
	
	Previous linear probing strategies, due to their heavy reliance on frozen encoder parameters $\theta$, were typically used to evaluate the quality of learned representations, thus limiting their further application. However, by treating $\theta$ as prior knowledge from large-scale representation learning, similar to human intuition in cognitive activities, we applied it to accurate physiological parameter prediction.
	
	Specifically, we harness the advantages of the fine-tuning approach, regarding the fine-tuning process as an reasoning process (RP) for specific downstream tasks, similar to human deliberation, integrating reasoning and prior results to achieve effective transfer to the task domain $\Phi$. Unlike conventional fine-tuning methods, the autonomous process (AP) serves as a prior constraint $H(\Phi|\theta)$, preventing domain bias due to poor quality of labeled data during fine-tuning. \cref{fig2} in the Downstream Task Adaptation section illustrates the DPT method.
	
	In order to integrate autonomous and inferential processes, we introduce the concept of Zero Decoder Layer (ZDL), where the neural layers in the decoder $g_{\delta}$ of the inference process RP are replaced by zero decoder layers. Assuming $h_{1}^{(l)}$ and $h_{2}^{(l)}$ are inputs of the encoder for RP and AP processes, the process of feature fusion and decoding in \cref{eq12} becomes as follows:
	
	\begin{equation}
		\left\{
		\begin{aligned}
			& a^{(l+1)} = ReLU\left[BatchNorm\left(Z_{\theta_0}(h_1^{(l)})\right)\right] \\
			& \qquad = ReLU\left[BatchNorm\left(W_0^{(l)}h_1^{(l)} + b_0^{(l)}\right)\right]
			\\
			& h_1^{(l+1)} = h_2^{(l)} + \psi^{(l+1)}\left(a^{(l+1)}\right)
		\end{aligned}
		\right.
		\label{eq14}
	\end{equation}
	
	It is worth noting that $Z_{\theta_0}(\cdot)$ is a decoder layer with weights $W_{0}^{(l)}$ and biases $b_{0}^{(l)}$ initialized to zero, only at the beginning of training when $h_{1}^{(l+1)}=h_{2}^{(l)}$, completely determined by prior intuition. Through subsequent fine-tuning training, specific inferential knowledge for downstream tasks will be integrated. This approach reduces the impact of harmful noise from downstream tasks on the inference process, while minimizing the risk of disrupting the initialization knowledge parameters of the inference process. This is a novel SSRL encoder transfer method, akin to human higher cognitive dual processes, integrating fine-tuning and linear probing through ZDL in the decoder, leveraging their respective strengths, yielding good results in multiple downstream tasks. Furthermore, DPT is a domain-agnostic transfer strategy that can be broadly applied to various self-supervised learning frameworks. Therefore, we also apply it to the temporal-spectrogram modeling in the proposed TS2TC framework.

	\begin{algorithm}[h]
		\caption{CTFGA with Dual-process Transfer (DPT)}
		\label{alg:DPT}
		\SetAlgoLined
		\KwIn{Pre-trained encoder $f_{\theta}$, 
			pre-trained decoder $g_{\delta}$,
			dilated signal $x_{A}^{\prime}(t)$ with ground truth labels $y$}
		\KwOut{Predicted PPG signal labels $\hat{y}$}
		\BlankLine
		
		\textbf{Initialize:} 
		$\theta_{new} \leftarrow \theta$, freeze $\theta$, freeze $\delta$, initialize $g_{new}$, $results\_queue = []$\;
		
		\For{epoch $t=1$ \KwTo $T$}{
			\For{each sample $(x^{\prime}(t), y)$}{
				\textbf{Autonomous Process (AP)}\\
				Compute $h_2^{(l)} = f_{\theta}(x_{A}^{\prime}(t))$\;
				\For{layer $l+1$ \KwTo $L$}{
					$b^{l+1}\in\delta$ and $W^{l+1}\in\delta $\;
					$a^{(l+1)} = W^{(l+1)}h_2^{(l)} + b^{(l+1)}$\;
					Update $h_2^{(l+1)} = \psi^{(l+1)}\bigl(a^{(l+1)}\bigr)$\;
					Append $h_2^{(l+1)}$ to $results\_queue$\;
				}
				
				\textbf{Reasoning Process (RP)}\\
				Compute $h_1^{(l)} = f_{\theta_{new}}(x_{A}^{\prime}(t))$\;
				\For{layer $l+1$ \KwTo $L$}{

					$h_2^{(l+1)} = results\_queue.front$\;

					Update $h_1^{(l+1)}$ using \cref{eq14}\;
					$results\_queue.pop$\;
				}
				Compute prediction $y' = g_{new}(h_1^{(L)})$
				
				\textbf{Loss and Backpropagation}\\
				Compute loss $\mathcal{L}_{MAE}(y', y)$\;
				Compute gradients:
				\begin{align*}
					\nabla_{\theta_{new},  W, b} \mathcal{L}_{MAE} &= \frac{\partial \mathcal{L}_{MAE}}{\partial y'} \cdot \frac{\partial y'}{\partial (f_{\theta_{new}}, g_{new})} \\
					&= \frac{\partial \mathcal{L}_{MAE}}{\partial y'} \cdot \frac{\partial (W f_{\theta_{new}}(x^{\prime}(t)) + b)}{\partial (f_{\theta_{new}}, g_{new})}
				\end{align*}
				Update parameters $\theta_{new}$ and $W, b$ using gradient descent:
				\[
				\theta_{new} = \theta_{new} - \eta \nabla_{\theta_{new}} \mathcal{L}_{MAE}, \quad  W, b =  W, b - \eta \nabla_{ W, b} \mathcal{L}_{MAE}
				\]
			}
		}
		
	\end{algorithm}

	\subsection{Spectrogram domain strategy}
	\label{Proposed TS2TC framework}
	
	\cref{CTFGA} has already described the components of TS2TC in the temporal domain, unlike the temporal domain, the input of the spectrogram domain is the spectrum map $x_{s}$ obtained through Short-time Fourier Transform (STFT):
	
	\begin{equation}
		X_m(f)=\sum_{n=0}^{N-1}x(n)w(n-mR)e^{-j\frac{2\pi}{N}fn},0\leq f\leq N-1
		\label{eq16}
	\end{equation}
	
	Here $x(n)$ represents the original PPG signal points of length $M\times T$ at the $n$th time point, where $f$ represents the frequency component, $N$ is the length of the DFT transformation interval, and $N\geq M\times T$. $w(\cdot)$ denotes a sliding window function of length $N$, by applying the $w(\cdot)$ function to the signal segment to obtain the complex-valued matrix $\mathbf{X}(f)=\{X_{1}(f),X_{2}(f),\cdots,X_{k}(f)\}$, the signal segment is centered at $mR$ in time, where $R$ is the window leap interval, thus knowing the number of overlapping sampling points $L=N-R$, where $k = \left \lfloor (M\times T-L)/(N-L) \right \rfloor $. Finally, by representing the amplitude in color, the matrix is projected onto the frequency domain and the temporal domain to obtain the spectrogram $x_{s}$. A schematic diagram of the spectrum can be seen in the Setup section of \cref{fig2}.
	
	Images are natural signals with a lot of spatial redundancy, we found that when $x_{s}$ randomly masks most patches, it encourages the model to regenerate the masked patches in pixel space, which will improve the prediction effectiveness of physiological parameters. Therefore, we adopt the Masked AutoEncoder (MAE) \citep{39he2022masked} as the backbone for modeling the spectrogram domain:
	
	\begin{equation}
		\begin{cases}x_s'=Mask(Pos(Patch(x_s)))\\z^{(l)}=Encoder(x_s')=ViTBlock^{l-1}(x_s')\end{cases}
		\label{eq17}
	\end{equation}
	
	And the strategy of the spectrogram domain is shown in \cref{alg:Spectrogram}. The encoder employs a Vision Transformer (ViT) \citep{40dosovitskiy2020image}, but only on visible and unmasked patches. During encoding, the image is first divided into patches and flattened into a 1-D sequence. After adding positional encoding, some patches are randomly masked to obtain $x_{s}^{\prime}$. It is processed by $l-1$ ViT blocks to produce encoded latent variables $z^{(l)}$. The decoder attempts to reconstruct $\hat{x}_s$ from $z^{(l)}$:
	
	\begin{equation}
		\hat{x}_s=Decoder(z^{(l)})=ViTBlock'\left(Pos\bigl(z^{(l)}\bigr)\right)
		\label{eq18}
	\end{equation}
	
	$ViTBlock'$ denotes multiple ViT blocks. The Self-Supervised Pretraining section in \cref{fig2} illustrates the pretraining process in the spectrogram domain. Our proposed Dual-Process Transfer (DPT) strategy is also applied to downstream tasks, as shown in the Downstream Task Adaptation section of \cref{fig2}. Suppose $z_{1}^{(l)}$ and $z_{2}^{(l)}$ are the inputs to the Representation Process (RP) and Adaptation Process (AP) encoders, respectively. The decoding process follows:
	
	\begin{equation}
		\begin{cases}a^{(l+1)}=ViTBlock_{unlock}^{(l+1)}\left(z_1^{(l)};\theta_1\right)\\z_1^{(l+1)}=ViTBlock_{lock}^{(l+1)}(z_2^{(l)};\theta_2)+a^{(l+1)}\end{cases}
		\label{eq19}
	\end{equation}
	
	$\theta_{1}$ and $\theta_{2}$ represent the parameters of the unfrozen and frozen ViT blocks in the decoder, respectively. Optimizing the network through Mean Absolute Error:
	
	\begin{equation}
		\mathcal{L}_{IR}=\frac{1}{n}\sum_{i=1}^n|x_{s_i}-\hat{x}_{s_i}|
		\label{eq19_}
	\end{equation}

	\begin{algorithm}[!h]
		\SetAlgoLined
		\caption{Spectrogram Domain Strategy}
		\label{alg:Spectrogram}
		\textbf{Phase 1: Pre-training with MAE}\;
		\KwIn{Spectrogram $x_s$ and $x_s'$}
		\KwOut{Trained neural network with encoder parameters $\theta_E$ and decoder parameters $\theta_2$, as well as latent variables $z^{(l)}$}
		$z^{(l)} \gets Encoder(x_s';\theta_E)$\;
		$\hat{x}_s \gets Decoder(z^{(l)};\theta_2)$\;
		Compute loss $\mathcal{L}_{IR}(\hat{x}_s, x_s)$\;
		Update parameters using gradient descent\;
		
		\BlankLine
		\textbf{Phase 2: Fine-tuning with DPT Strategy}\;
		\KwIn{$\theta_E$,$\theta_2$,$x_s'$,y}
		\KwOut{$z_1^{(l)}$, predicted PPG signal labels $\hat{y}$}
		\BlankLine
		$\theta_{unlock} \gets Encoder \text{ parameters } \theta_{E}$,$\theta_{lock} \gets \text{freeze } \theta_{E}$, freeze $\theta_{2}$, initialize $g_{new}$, $results\_queue = []$\;
		\For{epoch $t=1$ \KwTo $T$}{
			\For{each sample $(x_s', y)$}{
				\textbf{Autonomous Process (AP)}\\
				Compute $z_2^{(l)} \gets Encoder(x_s';\theta_{lock})$\;
				\For{layer $l+1$ \KwTo $L$}{
					Update $z_2^{(l+1)} = ViTBlock_{lock}^{(l+1)}(z_2^{(l)};\theta_2)$\;
					Append $z_2^{(l+1)}$ to $results\_queue$\;
				}
				\textbf{Reasoning Process (RP)}\\
				Compute $z_1^{(l)} \gets Encoder(x_s';\theta_{unlock})$\;
				\For{layer $l+1$ \KwTo $L$}{
					$z_2^{(l+1)} = results\_queue.front$\;
					$a^{(l+1)}=ViTBlock_{unlock}^{(l+1)}\left(z_1^{(l)};\theta_1\right)$\;
					Update $z_1^{(l+1)}=z_2^{(l+1)}+a^{(l+1)}$\;
					$results\_queue.pop$\;
				}
				Compute prediction  $y' = g_{new}(z_1^{(L)})$\;
				
				\textbf{Loss and Backpropagation}\;
				Compute loss $\mathcal{L}_{IR}(y', y)$\;
				Compute gradients:
				\[
				\nabla_{\theta_{unlock}, \theta_{1}, g_{new}} \mathcal{L}_{IR} = \frac{\partial \mathcal{L}_{IR}}{\partial \hat{x}_s} \cdot \frac{\partial \hat{x}_s}{\partial (Encoder, Decoder, g_{new})}
				\]
				Update parameters $\theta_{unlock}$, $\theta_{1}$ and $g_{new}$ using gradient descent:
				\[
				(\theta_{unlock}, \theta_{1}, g_{new}) = (\theta_{unlock}, \theta_{1}, g_{new}) - \eta \nabla_{\theta_{unlock}, \theta_{1}, g_{new}}
				\]
			}
		}
		
	\end{algorithm}

	Here, $n$ represents the total number of pixels in the spectrogram. Inspired by the BTSF method, we explicitly model the cross-domain dependencies between temporal and spectrogram pairs, enriching and differentiating representations through fusion and compression. Previous work overlooked the temporal-spectrogram relationship \citep{8kumar2023novel,21yang2022unsupervised}. Our proposed NBTSF method utilizes both temporal and frequency features, enhancing representation learning with finer granularity. This is the first application of the iterative bilinear method to unit time series. See \cref{alg:NBTSF} for details.
	
	\subsection{Novel bilinear temporal-spectrogram fusion (NBTSF)}
	\label{NBTSF}
	
	As mentioned in the CTFGA strategy, $h\in\mathbb{R}^{2E\times1}$ represents the fused latent space representation information in the past and future temporal domains, and the latent space representation $z^{(l)}\in\mathbb{R}^{D\times1}$ mentioned in the spectrogram domain strategy represents the latent space representation generated in the spectrogram domain by the encoder, where $D=E^{\prime}\times P$, $E'$ denotes the embedding dimension of Patch in the latent space, and $P$ denotes the number of unmasked Patches. We establish interactions between domain-specific features by inserting and decomposing the interaction factor $\sigma\in\mathbb{R}^{1\times1}$, aligning latent space representations across different domains:
	
	\begin{equation}
		F_{bilinear}=(W_1\times h)\times\sigma\times(W_2\times z)^T
		\label{eq20}
	\end{equation}
	
	Here, $W_{1}\in\mathbb{R}^{k\times2E}$ and $W_{2}\in\mathbb{R}^{k\times D}$ $(k=min(2E,D))$ are linear transformation matrices for generating minimal and aligned feature dimensions. We then expand the interaction factor $\sigma=UV^{T}$, where $U\in\mathbb{R}^{l\times1}$ and $V\in\mathbb{R}^{l\times1}$, to obtain bilinear feature representations:
	
	\begin{equation}
		\begin{split}
			F_{bilinear}&=(W_1\times h)\times U^T\times V\times(W_2\times z)^T\\
			&=\left((W_1\times h)\times U^T\right)\circ\left((W_2\times z)\times V^T\right)
		\end{split}
		\label{eq21}
	\end{equation}
	
	The symbol $\circ$ denotes the Hadamard product, which generates bilinear representations $F_{bilinear}\in\mathbb{R}^{k\times l}$ using two unbiased linear mappings, and $\times$ denotes matrix multiplication. This bilinear feature expresses fine-grained cross-domain affinities, refining the initial bilinear features adaptively through an iterative process, as shown below:
	
	\begin{equation}
		\begin{cases}S2T{:} h=BatchNorm\left(BiCasual\big(Conv1d(F_{bilinear})\big)\right)\\T2S{:} z=BatchNorm\left(Conv1d\big(BiCasual(F_{bilinear})\big)\right)\end{cases}
		\label{eq22}
	\end{equation}
	
	\begin{algorithm}[h]
		\SetAlgoLined
		\SetKwInOut{Input}{Input}
		\SetKwInOut{Output}{Output}
		
		\Input{Latent space representations $h\in\mathbb{R}^{2E\times1}$, $z^{(l)}\in\mathbb{R}^{D\times1}$}
		\Output{Enhanced bilinear feature representation $F_{enhance}$}
		
		\BlankLine
		Compute the initial bilinear feature representation $F_{bilinear}$ using \cref{eq21}\;
		
		\For{each iteration}{
			Apply S2T stage: $h=BatchNorm\left(BiCasual\big(Conv1d(F_{bilinear})\big)\right)$\;
			Apply T2S stage: $z=BatchNorm\left(Conv1d\big(BiCasual(F_{bilinear})\big)\right)$\;
			Update $F_{bilinear}$ using \cref{eq23}\;
		}
		
		Compute the final enhanced feature representation $F_{enhance}=W_1'\times h+W_2'\times z+F_{bilinear}$\;
		
		\caption{NBTSF Algorithm}
		\label{alg:NBTSF}
	\end{algorithm}

	Specifically, in the S2T stage, the Conv1d convolution block is first applied along the temporal-spectrogram axis to aggregate temporal-spectrogram information in each temporal feature. Subsequently, temporal features' dependencies are modeled by exchanging temporal-spectrogram-related information along the temporal axis using bidirectional causal convolution BiCasual \citep{41bai2018empirical} ; the T2S process is the reverse. Through the convolution-refined temporal and temporal-spectrogram features of S2T and T2S modules, more discriminative bilinear features are generated. By iteratively cycling through formulas \cref{eq21} and \cref{eq22}, a more cross-domain fusion and feature discriminative bilinear feature is collectively formed in a merged and compressed manner. The original temporal and temporal-spectrogram spectrum information are combined in the final bilinear representation:
	
	\begin{equation}
		F_{enhance}=W_1'\times h+W_2'\times z+F_{bilinear}
		\label{eq23}
	\end{equation}
	
\begin{figure}[h]
		\centering
		\includegraphics[width=50mm]{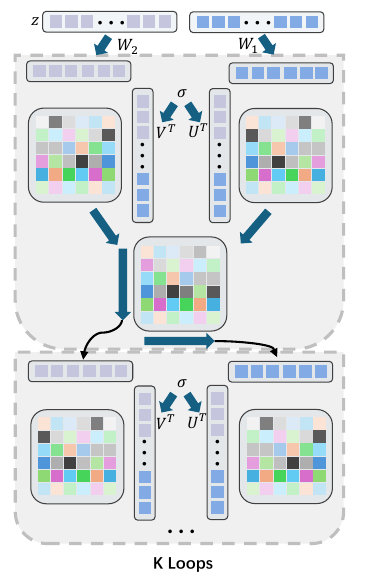}
		\caption{Novel Bilinear Temporal-Spectrogram Fusion (NBTSF)}
		\label{fig6}
	\end{figure}

	Here, both $W_{1}^{\prime}\in\mathbb{R}^{l\times2E}$ and $W_{2}^{\prime}\in\mathbb{R}^{l\times D}$ are linear transformation layers. \cref{fig6} illustrates the NBTSF process.
	
	\subsection{Time window relation generation (TWRG)}
	\label{Time window relation generation}
	
	A new generative pretext task for pretraining NBTSF, which contains a mixture of temporal and spectrogram domains in the mixed domain, is designed. The spectrogram domain provides complete temporal information $z$, while the temporal domain offers latent space embeddings $h_A=f_\theta(x_A^{\prime})$ and $h_C=f_\theta(x_C^{\prime})$. By confusing the time order of past information source $x_{A}^{\prime}(t)$ and future information source $x_{C}^{\prime}(t)$, positive samples $h_{positive}=\{h_{A},h_{C}\}$ and negative samples $h_{negative}=\{h_{C},h_{A}\}$ are generated. Then, $z$ containing temporal-spectrogram information along with randomly generated $h_{positive}$ or $h_{negative}$ are used as inputs to NBTSF to produce $F_{enhance}$. The final classification results $p(c_{i})=f_{\mu}(F_{enhance_{i}})$ are obtained through a classifier $f_{\mu}$ composed of MLP and Softmax, where $c_{i}\in\{positive,negative\}$. The loss function is defined using the cross-entropy function:
	
	\begin{equation}
		\mathcal{L}_{CLS}=-\frac{1}{|F_{enhance}|}\sum_{i=1}^{|F_{enhance}|}c_{i}\cdot log\left(p(c_{i})\right)+(1-c_{i})\cdot\log\bigl(1-p(c_{i})\bigr)
		\label{eq24}
	\end{equation}
	
	\subsection{Ternary fusion with ordinary least squares}
	\label{Ternary fusion with ordinary least squares}
	
	In many applications, understanding the reasons for a model's specific predictions and the accuracy of those predictions are equally important. However, achieving the highest accuracy on large modern datasets is typically done through complex models that can be hard to interpret even for experts. TS2TC leverages collaboration across multiple domains, thus necessitating an explanation of each domain's contribution to the overall TS2TC framework. In the Downstream Task Adaptation, we use the reference labels $y$ for physiological parameters as the gold standard, treating the predicted values $\hat{y}_{j}$ from each domain as attempts to interpret the features of the gold standard values. Interpreting the framework's final prediction $g({\hat{y}})$:
	
	\begin{equation}
		g(\hat{y})=\phi_0+\sum_{j=1}^M\phi_j\hat{y}_j\quad\text{s.t.}\min\|g(\hat{y})-y\|^2
		\label{eq25}
	\end{equation}
	
	This approach not only has the potential to enhance the accuracy of joint framework predictions but also allows for the explanation of each domain's contribution within the framework. For determining $\phi_{j}$, we utilize Ordinary Least Squares method:
	
	\begin{equation}
		\phi^T=(X^TX)^{-1}X^Ty
		\label{eq26}
	\end{equation}
	
	Here, $X\in\mathbb{R}^{m\times M}$, where $m$ denotes the number of training samples, $M=3$ represents the prediction dimensions, each from the temporal domain, spectrogram domain, and mixed domain, and $y\in\mathbb{R}^{m\times1}$ denotes the reference label values. $\phi\in\mathbb{R}^{M\times1}$ yields $\phi_{0}$ representing the interactive contribution, while $\phi_{j}$ signifies the independent contributions of each domain within the framework. Additionally, considering the R-squared of OLS, denoted as $R^{2}$, allows for quantifying the model's potential, i.e., $1-R^{2}$. \cref{Explanation of the contribution} and \cref{Ternary Fusion: Linear or Nonlinear} further delve into this method.
	
	\section{Experiments and results}
	\label{Experiments and results}
	
	\subsection{Datasets}
	\label{Datasets}
	
The Vital Signs DataBase (VitalDB) \citep{27lee2022vitaldb} contains unprocessed PPG records from 6177 surgical patients in the real world. Each record averages 2.8 million data points, with a sampling rate of 500 Hz. Due to missing data at the beginning and end of each record caused by the disconnection of the monitoring equipment, 40 non-overlapping and continuous sub-records of 7.2 seconds were uniformly and randomly selected from the middle 60\% of each record. This resulted in 247,080 records for experimentation. 90\% of the patients' PPG data were used for self-supervised pre-training, while the data from the remaining patients were used to validate the pre-training loss and adjust hyperparameters.
	
	The Bed-Based Ballistocardiography (BCG) database \citep{28carlson2020bed} contains PPG records from 40 individuals, including healthy individuals, from the real world. These records serve as a supplement to the observed data distribution for pre-training the model. Each record averages 0.4 million data points, with a sampling rate of 1 kHz. Sub-records were obtained by sliding segmentation with a step size of 0.8 seconds to obtain sufficient data. The BCG data contain filtered data, but we used the raw records without preprocessing. Similar to the VitalDB database, individuals were divided into pre-training and validation groups in a 9:1 ratio. All records used for pre-training and downstream tasks were resampled to 125 Hz.
	
	Eight databases from different sources were used to complete five downstream supervised tasks, ensuring that there was no patient overlap between these databases. The BIDMC database \citep{19zhang2024self,29goldberger2000physiobank} contains 7,949 records obtained from Physionet's BIDMC PPG and Respiration databases, which are extracted from the more extensive MIMIC II waveform database. The University of Sydney (UniSydney) database \citep{30mehrgardt2022pulse} contains PPG records from 22 healthy subjects performing three sports activities. To validate the framework's versatility, we used the BIDMC database for $\text{SpO}_{2}$, HR, and RR tasks, while the UniSydney database was used for $\text{SpO}_{2}$, HR, and BP tasks.
	
	For $\text{SpO}_{2}$ estimation, the University of Greifswald (UniGreifswald) database \citep{31blasing2022ecg} contains physiological measurements from 13 adult and healthy subjects during a standardized experimental setup. The database for $\text{SpO}_{2}$ has mean and standard deviation values of 95.96 and 29.98, respectively. For HR estimation, the IEEEPPG dataset \citep{32zhang2014troika} contains 3,096 records obtained from the IEEE Signal Processing Cup 2015. The mean and standard deviation of HR values used as labels across all experimental databases are 94.15 and 15.30, respectively. For RR estimation, the Pulse Wave Database (PWDB) \citep{33charlton2019modeling} contains simulated PPG records from 4,374 healthy adults aged 25-75. The CapnoBase TBME RR (CapnoBase) benchmark database \citep{34karlen2013multiparameter} includes 42 8-minute records from pediatric and adult surgeries and anesthesia. The mean and standard deviation of RR values used as labels across all experimental databases are 17.96 and 5.37, respectively. For BP estimation, the PPGBP database \citep{35liang2018new} contains a total of 613 short PPG records of 2.1 seconds each from 219 subjects, making it the smallest set in terms of segment count but relatively larger in terms of subject count. We unified the input length by replicating the short records four times. The Non-invasive Blood Pressure Estimation (NBPE) dataset \citep{36esmaili2017nonlinear} provides a collection of vital signals and reference blood pressure values from 26 subjects. Across all experimental databases, the mean and standard deviation of systolic blood pressure used as labels are 121.87 and 15.18, respectively. For diastolic blood pressure, the mean and standard deviation are 76.44 and 8.31, respectively. The time-series windows were obtained by sliding with a step size of 0.8 seconds, with 70\% of individual records assigned to the training set, 10\% to the validation set, and the remaining data to the test set.
	
	Additionally, it is noteworthy that the data partitioning method in this work adheres to the inter-subject (or patient-based) paradigm. \Cref{tab1} summarizes the information of the datasets used in the experiments.
	
	\begin{table}[h]
		\centering
		\caption{Statistical distribution of the PPG databases.}
		\resizebox{\linewidth}{!}{
			\begin{tabular}{cp{4.7em}cccp{4.19em}p{4.19em}}
				\toprule\toprule
				\multicolumn{1}{c}{Database} & \multicolumn{3}{c}{Pretraining} & \multicolumn{3}{c}{Validation} \\
				\midrule
				\multicolumn{1}{c}{VitalDB} & \multicolumn{3}{c}{222360} & \multicolumn{3}{c}{24720} \\
				\multicolumn{1}{c}{BCG} & \multicolumn{3}{c}{24768} & \multicolumn{3}{c}{2752} \\
				\midrule
				\multicolumn{1}{c}{Class} & Database & \multicolumn{1}{c}{Training} & \multicolumn{1}{c}{Validation} & \multicolumn{1}{c}{Test} & Length & Sampling Rate (Hz) \\
				\midrule
				\multicolumn{1}{c}{\multirow{4}[4]{*}{$\text{SpO}_{2}$}} & BIDMC & 178057 & 25436 & 50875 & \multicolumn{1}{c}{4000} & \multicolumn{1}{c}{125} \\
				& UniGreifswald & 29976 & 3138  & 9566  & \multicolumn{1}{c}{1316000} & \multicolumn{1}{c}{512} \\
				& UniSydney & 27854 & 3649  & 8479  & \multicolumn{1}{c}{245000} & \multicolumn{1}{c}{500} \\
				\cmidrule{2-7}          & \textbf{Total} & \textbf{235887} & \textbf{32223} & \textbf{68920} & \multicolumn{1}{c}{-}     & \multicolumn{1}{c}{-} \\
				\midrule
				\multicolumn{1}{c}{\multirow{4}[4]{*}{HR}} & BIDMC & 178057 & 25436 & 50875 & \multicolumn{1}{c}{4000} & \multicolumn{1}{c}{125} \\
				& IEEEPPG & 2167  & 309   & 620   & \multicolumn{1}{c}{1000} & \multicolumn{1}{c}{125} \\
				& UniSydney & 27854 & 3649  & 8479  & \multicolumn{1}{c}{245000} & \multicolumn{1}{c}{500} \\
				\cmidrule{2-7}          & \textbf{Total} & \textbf{208078} & \textbf{29394} & \textbf{59974} & \multicolumn{1}{c}{-}     & \multicolumn{1}{c}{-} \\
				\midrule
				\multicolumn{1}{c}{\multirow{4}[4]{*}{RR}} & BIDMC & 178057 & 25436 & 50875 & \multicolumn{1}{c}{4000} & \multicolumn{1}{c}{125} \\
				& PWDB  & 34036 & 4826  & 9906  & \multicolumn{1}{c}{105000} & \multicolumn{1}{c}{500} \\
				& CapnoBase & 17168 & 2368  & 5328  & \multicolumn{1}{c}{144000} & \multicolumn{1}{c}{300} \\
				\cmidrule{2-7}          & \textbf{Total} & \textbf{229261} & \textbf{32630} & \textbf{66109} & \multicolumn{1}{c}{-}     & \multicolumn{1}{c}{-} \\
				\midrule
				\multicolumn{1}{c}{\multirow{4}[8]{*}{BP}} & UniSydney & 27854 & 3649  & 8479  & \multicolumn{1}{c}{245000} & \multicolumn{1}{c}{500} \\
				\cmidrule{2-7}          & PPGBP & 918   & 126   & 270   & \multicolumn{1}{c}{8400} & \multicolumn{1}{c}{1000} \\
				\cmidrule{2-7}          & NBPE  & 7952  & 867   & 3190  & \multicolumn{1}{c}{376000} & \multicolumn{1}{c}{1000} \\
				\cmidrule{2-7}          & \textbf{Total} & \textbf{36724} & \textbf{4642} & \textbf{11939} & \multicolumn{1}{c}{-}     & \multicolumn{1}{c}{-} \\
				\bottomrule
			\end{tabular}%
		}
		\label{tab1}%
	\end{table}%
	
	\subsection{Data preprocessing}
	\label{Data preprocessing}
	
	The process of PPG preprocessing can be summarized in two steps:

	\textbf{Resampling.} Due to different devices and settings, each database has a different sampling rate $L$, so it needs to be uniformly resampled to $M(\mathrm{Hz})\times T(s)$ data points. This study chose a sampling rate of $M = 125$, which not only reduces computational costs while retaining sufficient information but is also the minimum sampling rate for most existing PPG databases. \cref{Exploration of CTFGA's best γ and T} explores the effect of different sampling times $T$ on the results with a step size of 0.8 seconds, where 0.8 seconds is the duration of a single cardiac cycle. Based on the results of the exploration, $T = 5.6$ was chosen.
	
	\textbf{Normalization.} Each record is standardized using z-score normalization. Let $x(t)$ represent the PPG signal. The z-score can be formulated as:
	\begin{equation}
		x(t)=\frac{x(t)-\mu}{\delta} 
		\label{eq27}
	\end{equation}
	where $\mu$ and $\delta$ are the mean and standard deviation of $x(t)$, respectively; the purpose of the z-score is to eliminate amplitude differences caused by different data sources.
	
	\subsection{Experimental settings}
	\label{Experimental settings}
	
	The Adagrad optimizer was used for model training with a learning rate of 0.001 and a batch size of 32. The parameter $\gamma$ in CTFGA was set to 0.4. For both pre-training and downstream task stages, the models were trained for 100 and 300 epochs, respectively. Mean Absolute Error (MAE) and Root Mean Square Error (RMSE) were used to evaluate the regression performance, while Mean Absolute Percentage Error (MAPE) was used to indicate the relative prediction accuracy of different physiological parameters. The results of MAE and RMSE are reported as the (Mean $\pm$ Std) over ten random and independent experiments. The settings for the comparative generative self-supervised representation learning followed these parameters. All models were implemented using the PyTorch framework, and experiments were conducted using 2 NVIDIA RTX 4090 GPUs with 24 GB of memory.
	
	\subsection{Results: CTFGA}
	\label{Results: CTFGA}
	
\begin{figure*}[h]
		\centering
		\includegraphics[width=180mm]{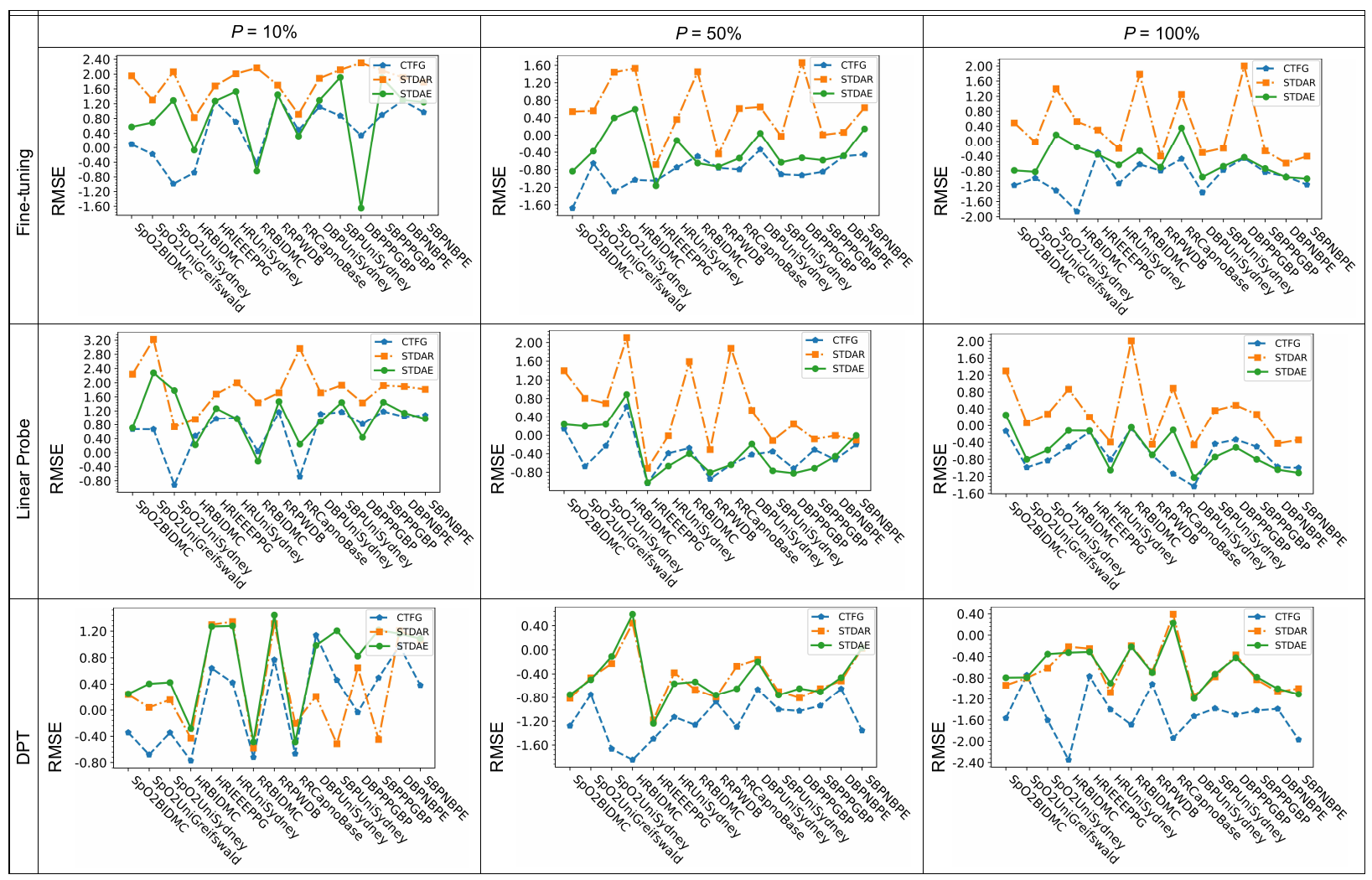}
		\caption{Transfer results under different pre-training methods and data amounts. (Standardized)}
		\label{fig7}
	\end{figure*}

\begin{figure*}[h]
		\centering
		\includegraphics[width=185mm]{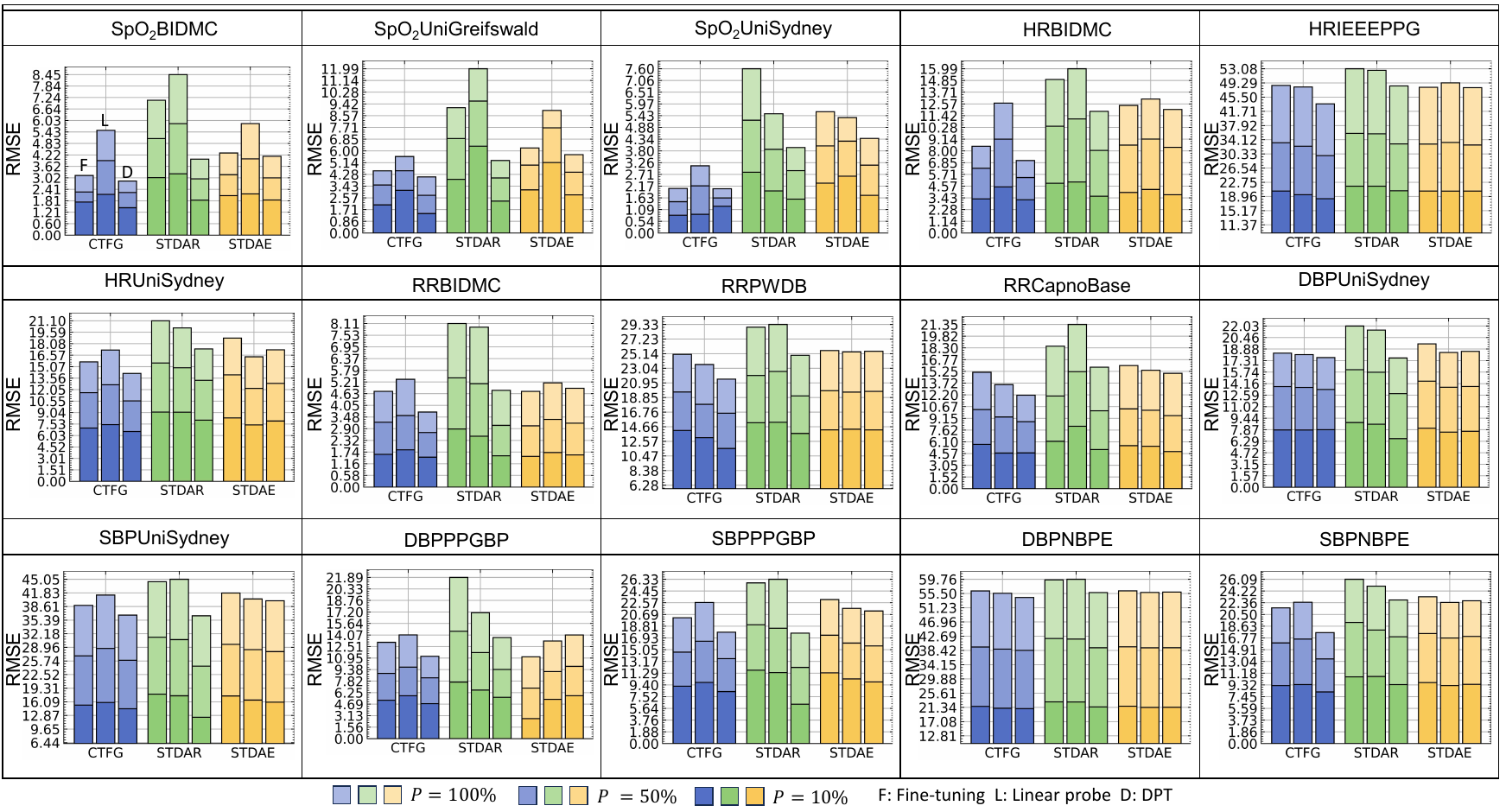}
		\caption{Pre-training results under different transfer methods and data amounts ($P$: proportion of training data, SBP: systolic blood pressure, DBP: diastolic blood pressure).}
		\label{fig8}
	\end{figure*}

	\cref{fig7} illustrates the relative superiority of the CTFGA method under different encoder transfer task settings, where standardization is used to weaken the differences between data domains to highlight the superiority of the task. The non-standardized values can be seen in \cref{fig8}. To compare the effectiveness of the CTFGA strategy on the same scale, we applied standard autoregressive (STDAR) and standard autoencoder (STDAE) methods on the temporal domain network architecture proposed in this paper, including the use of VMD and multi-stage derivative staged injection for feature expansion and injection. All methods followed the same pre-training and fine-tuning approach on the same dataset, thus ignoring differences in the underlying architecture and focusing on the effectiveness of different generative self-supervised pretext tasks.
	
	Fine-tuning demonstrates the effectiveness of initializing weights with various pretext tasks in the transfer process of downstream tasks. Compared to learning features from partial single information sources to infer the remaining unknown information, especially the external spatial uncertainties caused by body movement, we found that the complete instance-level sequence autoencoder is significantly better than autoregressive due to the emphasis on feature-level rather than strictly point-level contextual temporal information in physiological parameter estimation. This is also why this framework uses ordinary 1-D convolution without considering time leakage and adopts causal convolution with strict time constraints. However, instance-level reconstruction can still implicitly model contextual temporal information. Through the pretext task modeling process of generating coarse-grained temporal relationships and sequence reconstruction with multiple information sources, CTFGA is significantly better than autoregressive and autoencoder. Therefore, explicit modeling of contextual temporal relationships at the fragment level of PPG sequences is more effective than implicit modelling.
	
	The linear probe method can observe the quality of the representations learned by different generative pretext tasks. We found that both fine-tuning and linear probes demonstrate consistency in predicting trends on different data domains (i.e., databases) of the same physiological parameters and different distributions within the same data domain (i.e., different data amounts). CTFGA performs more prominently in linear probes, indicating that it has effectively learned robust generalized features. In DPT, due to fully considering the effectiveness of transfer learning of learned generalized features in pre-training and downstream tasks, the performance of autoregressive is improved, narrowing the difference between autoregressive and autoencoder tasks. The representations learned by CTFGA benefit from this effectiveness, showing a significant advantage over the same type of pretext tasks at 50\% and 100\% of the training data proportion.
	
	\subsection{Results: DPT}
	\label{Results: DPT}
	
	The superiority of the combination of DPT and CTFGA is already seen in \cref{fig7}, and \cref{fig8} shows explicitly the advantages of DPT in various pretext tasks under different encoder transfer strategies. The first column of each color group represents the fine-tuning method transfer, the second column represents the linear probe transfer method, and the third column uses our proposed DPT method, with different shades of the same series representing different data amounts. Consistent with the conclusion in \cref{Results: CTFGA}, the DPT transfer strategy is most effective for autoregressive, which is based on predicting inference, because the autonomous process AP of DPT serves as a prior constraint $H(\Phi|\theta)$, which can effectively avoid domain bias due to differences in the quality of annotated data during fine-tuning. At the same time, it is observed that with 50\% of the training data, both fine-tuning and linear probes show worse performance than with 10\% of the training data. This is because the harmful noise of the downstream task affects the inference process, increasing the risk of disrupting the initialization of knowledge parameters for inference. At the same time, the effectiveness of DPT decreases steadily with the increase in data amounts, reflecting the robustness brought by ZDL.

		\begin{table}[h]
		\centering
		\caption{Performance of components in each composition of TS2TC. (bold represents the best results in different domains)}
		\resizebox{\linewidth}{!}{
			\begin{tabular}{clcccccc}
				\toprule\toprule
				& Component & \multicolumn{2}{c}{Temporal} & \multicolumn{2}{c}{Spectrogram} & \multicolumn{2}{c}{Mixed} \\
				\midrule
				& Database & MAE   & MAPE  & MAE   & MAPE  & MAE   & MAPE \\
				\midrule
				\multirow{3}[2]{*}{$\text{SpO}_{2}$ (\%)} & BIDMC & 1.007±0.0485 & 0.010  & 2.297±0.218 & 0.024  & \textbf{0.996±0.085} & \textbf{0.010 } \\
				& UniGreifswald & 0.947±0.0215 & 0.010  & \textbf{0.463±0.041} & \textbf{0.010 } & 0.514±0.042 & 0.011  \\
				& UniSydney & 0.467±0.0175 & 0.005  & \textbf{0.273±0.011} & \textbf{0.004 } & 0.427±0.024 & 0.006  \\
				\midrule
				\multirow{3}[2]{*}{HR (bpm)} & BIDMC & \textbf{1.879±0.074} & \textbf{0.022 } & 10.41±1.105 & 0.104  & 8.001±0.687 & 0.080  \\
				& IEEEPPG & \textbf{15.293±0.2405} & \textbf{0.149 } & 22.525±0.669 & 0.239  & 16.439±1.482 & 0.174  \\
				& UniSydney & 4.93±0.083 & 0.067  & \textbf{4.265±0.184} & \textbf{0.058 } & 4.662±0.385 & 0.064  \\
				\midrule
				\multirow{3}[2]{*}{RR (bpm)} & BIDMC & \textbf{0.961±0.0045} & \textbf{0.075 } & 3.143±0.941 & 0.297  & 2.304±0.233 & 0.217  \\
				& PWDB  & 7.925±0.318 & 0.486  & 6.489±0.337 & 0.454  & \textbf{5.281±0.315} & \textbf{0.370 } \\
				& CapnoBase & 3.258±0.108 & 0.206  & 6.225±1.514 & 0.387  & \textbf{2.388±0.162} & \textbf{0.148 } \\
				\midrule
				\multirow{3}[2]{*}{DBP (mmHg)} & UniSydney & 6.21±0.1425 & 0.081  & 6.64±0.323 & 0.031  & \textbf{6.169±0.378} & \textbf{0.028 } \\
				& PPGBP & 11.28±0.0735 & 0.158  & \textbf{9.066±0.863} & \textbf{0.131 } & 10.01±1.059 & 0.144  \\
				& NBPE  & \textbf{3.757±0.0475} & \textbf{0.053 } & 6.786±0.605 & 0.125  & 4.121±0.14 & 0.076  \\
				\midrule
				\multirow{3}[2]{*}{SBP (mmHg)} & UniSydney & 6.006±0.278 & 0.052  & 8.908±0.565 & 0.027  & \textbf{5.958±0.605} & \textbf{0.029 } \\
				& PPGBP & 16.639±0.737 & 0.142  & \textbf{13.362±1.216} & \textbf{0.123 } & 14.893±1.482 & 0.137  \\
				& NBPE  & \textbf{6.065±0.17} & \textbf{0.041 } & 18.551±0.7 & 0.144  & 23.923±0.913 & 0.185  \\
				\bottomrule
			\end{tabular}%
		}
		\label{tab2}%
	\end{table}%
	
	We also found that there was no significant difference between the three encoder transfer strategies in the case of autoencoders because the representations learned by the autoencoder as a pretext task lack diversity. Although the performance of the autoencoder is better than autoregressive without DPT intervention and even comparable to our results (\cref{fig8}), the representation of latent space is limited by learning monotonic mappings. Therefore, estimating uncertainty, as in autoregressive and DPT, is equally essential for generative SSRL. Preserving the learning of uncertainty can effectively further optimize network learning. With the increase of data amounts, the inference process RP through DPT can gradually transform this uncertainty into specific deterministic inference for specific downstream tasks, thereby improving the model performance.
	
	\subsection{Results: TS2TC}
	\label{Results: TS2TC}
	As shown in \cref{fig2}, the TS2TC framework for downstream tasks mainly includes the temporal domain, spectrogram domain based on CTFGA and DPT strategies, and the mixed domain based on the novel Bilinear Temporal-Spectrogram Fusion (NBTSF) strategy. Finally, the Ordinary Least Squares (OLS) method is used for aggregation to obtain the final output of the TS2TC framework. \Cref{tab2} presents the performance of TS2TC components in predicting various physiological parameters under conditions of limited data samples (10\% training data proportion).
	
	The original temporal domain achieves the best results in most tasks. However, the spectrogram domain performs better on UniGreifswald, PPGBP, and PWDB databases, indicating that the spectrogram domain can provide complementary information to the joint components. Furthermore, PWDB, BIDMC for $\text{SpO}_{2}$, and UniSydney for BP also show significant advantages in the mixed domain. Since the actual effectiveness of each component cannot be predicted in practical prediction, the final OLS balances the components in the three domains. The aggregated results are shown in the TS2TC column of \cref{tab3}. On average, the TS2TC outperforms existing best estimation methods by 2.49\% in RMSE with 10\% training data proportion while considering the predictions as interpretations of the ground truth labels.
	
\begin{figure}[h]
		\centering
		\begin{subfigure}{0.34\textwidth}
			\centering
			\includegraphics[width=\textwidth]{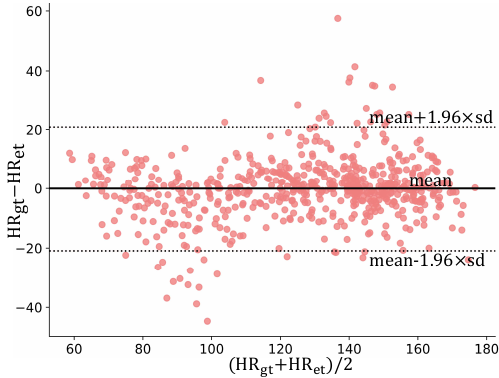}
			\caption{}
			\label{fig9a}
		\end{subfigure}
		\hspace{1cm}
		\begin{subfigure}{0.34\textwidth}
			\centering
			\includegraphics[width=\textwidth]{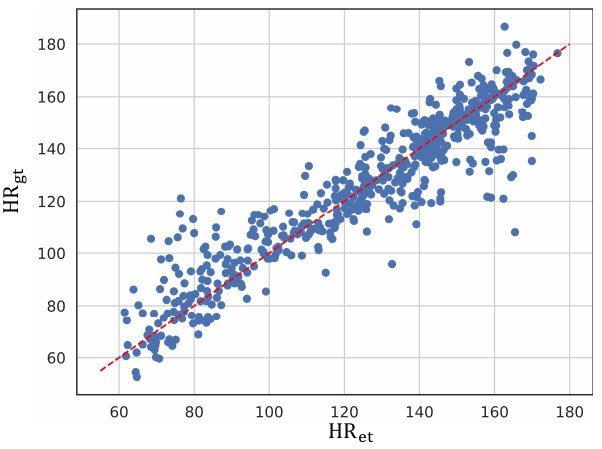}
			\caption{}
			\label{fig9b}
		\end{subfigure}
		\caption{The results of the TS2TC method on IEEEPPG (P=10\%).}
		\label{fig9}
	\end{figure}

	\cref{fig9} shows the TS2TC results on the IEEEPPG database, a widely used database for heart rate prediction, with only 10\% of the data participating in downstream task training. $\text{HR}_{\text{gt}}$ represents the ground truth, i.e., the actual label value, and $\text{HR}_{\text{et}}$ represents the estimated value. The Bland-Altman plot in \cref{fig9a} demonstrates the consistency advantage of TS2TC in physiological parameter estimation compared to the gold standard measurement. \cref{fig9b} shows the high correlation between the proposed method and the standard method.

	\begin{table*}[bp]
		\centering
		\caption{Comparison of results of TS2TC framework under sample scarcity situation (MAE and MAPE $\pm$ Std), where bold and underline represent the best performance among the comparison methods, and underline only represents the second best performance.}
		\resizebox{\linewidth}{!}{
			\begin{tabular}{cc|ccc|ccc|ccc|ccc|cccr}
				\toprule\toprule
				&       & \multicolumn{3}{c|}{$\text{SpO}_{2}$ (\%)} & \multicolumn{3}{c|}{HR (bpm)} & \multicolumn{3}{c|}{RR (bpm)} & \multicolumn{3}{c|}{DBP (mmHg)} & \multicolumn{3}{c}{SBP (mmHg)} & \multicolumn{1}{c}{\multirow{2}[4]{*}{Average}} \\
				\cmidrule{1-17}    Models & Metric & BIDMC & UniGreifswald & UniSydney & BIDMC & IEEEPPG & UniSydney & BIDMC & PWDB  & CapnoBase & UniSydney & PPGBP & NBPE  & UniSydney & PPGBP & NBPE  &  \\
				\midrule
				\multirow{2}[2]{*}{TFTC} & MAE   & \underline{\textbf{1.217±0.103}} & \underline{\textbf{0.608±0.051}} & \underline{\textbf{0.428±0.035}} & 4.094±0.172 & \underline{\textbf{16.305±0.613}} & \underline{\textbf{3.950±0.236}} & 1.445±0.095 & \underline{\textbf{4.832±0.171}} & \underline{\textbf{4.707±0.141}} & \underline{\textbf{8.571±0.367}} & 9.576±0.303 &\underline{ \textbf{3.426±0.149}} & \underline{\textbf{5.504±0.550}} & \underline{\textbf{12.836±1.347}} & 9.352±0.827 & \underline{\textbf{5.790 }} \\
				& MAPE  & 0.022 & 0.006 & 0.006 & 0.041 & 0.173 & 0.053 & 0.136 & 0.338 & 0.292 & 0.035 & 0.088 & 0.033 & 0.027 & 0.178 & 0.072 & 0.100  \\
				\midrule
				\multirow{2}[2]{*}{PatchTST} & MAE   & 2.537±0.101 & 4.749±0.390 & 11.261±0.472 & 3.097±0.300 & 27.797±2.348 & 8.180±0.550 & 2.029±0.057 & 14.016±0.598 & \underline{6.033±0.365} & 57.325±5.869 & 19.729±0.956 & 6.230±0.551 & 9.300±0.871 & 17.437±1.014 & 14.582±0.884 & 13.620  \\
				& MAPE  & 0.027 & 0.049 & 0.115 & 0.035 & 0.262 & 0.108 & 0.159 & 0.909 & 0.364 & 0.757 & 0.281 & 0.089 & 0.080  & 0.143 & 0.106 & 0.232  \\
				\midrule
				\multirow{2}[2]{*}{TSSequencer} & MAE   & 2.456±0.229 & 0.907±0.046 & \underline{0.532±0.023} & \underline{\textbf{1.834±0.147}} & 24.746±1.230 & 7.839±0.398 & 1.817±0.187 & 14.018±0.898 & 6.107±0.444 & 56.396±5.521 & \underline{9.473±0.980} & 5.985±0.534 & 9.222±0.941 & 17.526±1.138 & 12.829±0.625 & \underline{11.446}  \\
				& MAPE  & 0.026 & 0.009 & 0.005 & 0.022 & 0.230  & 0.106 & 0.142 & 0.868 & 0.395 & 0.743 & 0.132 & 0.085 & 0.079 & 0.145 & 0.093 & 0.205  \\
				\midrule
				\multirow{2}[2]{*}{TSPerceiver} & MAE   & 2.419±1.832 & 1.039±0.050 & 88.623±9.587 & 9.332±0.713 & 23.988±2.219 & 7.943±0.283 & 2.158±0.112 & 14.775±0.485 & 18.630±0.847 & 75.263±4.749 & 31.553±1.632 & 5.383±0.160 & 9.132±0.713 & 49.670±4.367 & 12.374±0.717 & 23.485  \\
				& MAPE  & 0.026 & 0.011 & 0.912 & 0.106 & 0.233 & 0.107 & 0.183 & 0.629 & 1.090  & 0.997 & 0.473 & 0.078 & 0.078 & 0.424 & 0.090  & 0.362  \\
				\midrule
				\multirow{2}[2]{*}{XCM} & MAE   & 2.110±0.204 & 1.399±0.037 & 86.263±5.697 & \underline{1.860±0.064} & 45.333±2.445 & 6.269±0.306 & 1.513±0.078 & 11.676±0.376 & 17.145±1.783 & 73.767±6.970 & 13.886±1.093 & 5.561±0.437 & 8.861±0.588 & 26.489±1.133 & 10.926±0.514 & 20.871  \\
				& MAPE  & 0.022 & 0.014 & 0.887 & 0.022 & 0.348 & 0.087 & 0.114 & 0.662 & 0.954 & 0.977 & 0.183 & 0.080  & 0.076 & 0.209 & 0.080  & 0.314  \\
				\midrule
				\multirow{2}[2]{*}{MultiRocket} & MAE   & \underline{1.811±0.051} & 3.751±0.295 & 7.128±0.234 & 2.094±0.088 & 18.058±1.949 & \underline{4.670±0.187} & 1.123±0.028 & \underline{6.794±1.935} & 17.880±0.602 & 74.835±3.598 & 9.926±0.374 & 5.032±0.185 & \underline{5.930±0.348} & 18.944±0.708 & \underline{7.691±0.683} & 12.378  \\
				& MAPE  & 0.019 & 0.038 & 0.073 & 0.024 & 0.165 & 0.065 & 0.084 & 0.352 & 1.114 & 0.998 & 0.139 & 0.072 & 0.051 & 0.156 & 0.056 & 0.227  \\
				\midrule
				\multirow{2}[2]{*}{TSiTransformer } & MAE   & 2.434±2.568 & \underline{0.886±0.028} & 0.805±0.054 & 10.055±0.787 & 24.016±0.944 & 7.860±0.416 & 2.277±0.144 & 14.012±1.351 & 7.935±0.627 & 59.744±2.470 & \textbf{\underline{9.377±0.456}} & 5.986±0.608 & 9.251±0.531 & 38.792±3.918 & 12.834±1.356 & 13.751  \\
				& MAPE  & 0.026 & 0.009 & 0.008 & 0.115 & 0.230  & 0.106 & 0.193 & 0.864 & 0.359 & 0.788 & 0.127 & 0.085 & 0.080  & 0.284 & 0.094 & 0.225  \\
				\midrule
				\multirow{2}[2]{*}{TST} & MAE   & 4.832±0.363 & 9.927±0.876 & 4.418±0.192 & 13.524±0.966 & 25.669±0.736 & 17.075±0.689 & 2.423±0.110 & 18.608±1.894 & 6.821±2.056 & \underline{13.384±1.433} & 26.087±1.382 & 7.179±0.449 & 9.606±0.545 & 23.738±2.370 & 16.436±1.443 & 13.315  \\
				& MAPE  & 0.050  & 0.101 & 0.045 & 0.145 & 0.234 & 0.214 & 0.196 & 0.656 & 0.379 & 0.171 & 0.355 & 0.100 & 0.083 & 0.187 & 0.122 & 0.203  \\
				\midrule
				\multirow{2}[2]{*}{OmniScale} & MAE   & 1.825±0.164 & 1.322±0.108 & 85.204±6.287 & 2.577±0.214 & \underline{16.694±0.636} & 5.439±0.250 & \underline{1.097±0.087} & 12.322±0.761 & 17.913±1.079 & 74.355±6.905 & 35.307±3.543 & \underline{3.958±0.136} & 7.294±0.274 & 18.303±0.767 & \underline{\textbf{7.048±0.363}} & 19.377  \\
				& MAPE  & 0.019 & 0.014 & 0.877 & 0.019 & 0.153 & 0.076 & 0.087 & 0.539 & 0.967 & 0.985 & 0.473 & 0.056 & 0.063 & 0.145 & 0.051 & 0.302  \\
				\midrule
				\multirow{2}[2]{*}{InceptionTime} & MAE   & 2.103±0.222 & 2.352±0.217 & 60.879±1.782 & 2.054±0.074 & 18.885±0.906 & 5.744±0.306 & \underline{\textbf{1.049±0.094}} & 12.412±1.341 & 15.763±1.532 & 71.412±1.969 & 10.922±0.417 & 4.687±0.306 & 7.082±0.665 & \underline{\textbf{17.020±1.440}} & 8.841±0.500 & 16.080  \\
				& MAPE  & 0.022 & 0.024 & 0.627 & 0.024 & 0.181 & 0.080  & 0.084 & 0.762 & 0.836 & 0.945 & 0.151 & 0.062 & 0.061 & 0.139 & 0.065 & 0.271  \\
				\bottomrule
			\end{tabular}%
		}
		\label{tab3}%
	\end{table*}%
	
\section{Discussion}
	\label{Discussion}
	\subsection{Comparison with existing studies}
	\label{Comparison with existing studies}
	
	Research on generative self-supervised representation learning for estimating physiological parameters from PPG focuses on the temporal domain, which is also the central aspect of our study. In addition to comparing STDAR and STDAE for self-supervised learning, we also contrasted them with other generative models. These include the transformer-based TST framework \citep{43zerveas2021transformer}, PatchTST \citep{44nie2022time}, which explicitly preserves local semantic information in embeddings, TSiTransformer, a ViT-based time-series implementation \citep{45dosovitskiy2020image}, TSSequencer, a time-series implementation of Sequencer \citep{46tatsunami2022sequencer} challenging the Transformer with CNNs, and TSPerceiver, a time-series implementation of Perceiver IO \citep{47jaegle2021perceiver} using attention non-uniformly. Additionally, we compared them with supervised time-series models, including MultiRocket \citep{48tan2022multirocket}, which improves feature diversity by adding multiple pooling operators and transformations, as well as XCM \citep{49fauvel2021xcm}, OmniScale \citep{50tang2021omni}, and InceptionTime \citep{51ismail2020inceptiontime}.
	\begin{table*}[h]
		\centering
		\caption{Comparison of results based on CTFGA and DPT (RMSE±Std and MAPE), where bold indicates the best performance of the method, and underline indicates the second-best performance method.}
		\resizebox{\linewidth}{!}{
			\renewcommand{\arraystretch}{1.62}
			\begin{tabular}{ccc|cccccccccccccccccccc}
				\toprule\toprule
				\multicolumn{3}{c|}{Models} & \multicolumn{2}{c}{CTFG-DPT} & \multicolumn{2}{c}{PatchTST} & \multicolumn{2}{c}{TSSequencer} & \multicolumn{2}{c}{TSPerceiver} & \multicolumn{2}{c}{XCM} & \multicolumn{2}{c}{MultiRocket} & \multicolumn{2}{c}{TSiTransformer } & \multicolumn{2}{c}{TST} & \multicolumn{2}{c}{OmniScale} & \multicolumn{2}{c}{InceptionTime} \\
				\midrule
				\multicolumn{3}{c|}{Metric} & RMSE  & MAPE  & RMSE  & MAPE  & RMSE  & MAPE  & RMSE  & MAPE  & RMSE  & MAPE  & RMSE  & MAPE  & RMSE  & MAPE  & RMSE  & MAPE  & RMSE  & MAPE  & RMSE  & MAPE \\
				\midrule
				\multirow{9}[6]{*}{$\text{SpO}_{2}$ (\%)} & \multirow{3}[2]{*}{\begin{sideways}BIDMC\end{sideways}} & 10\%  & \textbf{1.444±0.156} & \textbf{0.010 } & 3.384±0.142 & 0.027  & 3.276±0.156 & 0.026  & 3.245±0.251 & 0.026  & 2.814±0.243 & 0.022  & \underline{2.445±0.149} &\underline{ 0.019}  & 3.277±0.115 & 0.026  & 5.506±0.268 & 0.050  & 2.615±0.277 & 0.019  & 2.963±0.188 & 0.022  \\
				&       & 50\%  & \textbf{0.800±0.065} & \textbf{0.005 } & 3.253±0.263 & 0.025  & 3.277±0.126 & 0.026  & 3.233±0.155 & 0.025  & 2.053±0.122 & 0.015  & 1.236±0.078 & 0.009  & 3.276±0.328 & 0.026  & 3.839±0.164 & 0.027  & \underline{1.225±0.119} & \underline{0.009}  & 1.506±0.115 & 0.011  \\
				&       & 100\% & \textbf{0.603±0.061} & \textbf{0.004 } & 3.275±0.246 & 0.026  & 3.276±0.285 & 0.026  & 3.202±0.239 & 0.025  & 1.514±0.077 & 0.011  & 1.138±0.120 & 0.008  & 3.276±0.217 & 0.026  & 3.265±0.512 & 0.026  & \underline{1.006±0.099} & \underline{0.007}  & 1.100±0.092 & 0.008  \\
				\cmidrule{2-23}          & \multirow{3}[2]{*}{\begin{sideways}UniGreifswald\end{sideways}} & 10\%  & 1.422±0.102 & 0.010  & 19.515±2.024 & 0.049  & \underline{1.328±0.098} & \underline{0.009}  & 1.590±0.092 & 0.011  & 2.058±0.062 & 0.014  & 5.373±0.307 & 0.038  & \textbf{1.306±0.092} & \textbf{0.009 } & 10.367±0.794 & 0.101  & 1.647±0.165 & 0.014  & 8.572±0.449 & 0.024  \\
				&       & 50\%  & 1.333±0.110 & 0.009  & 19.497±0.866 & 0.046  & 1.328±0.035 & 0.009  & 1.330±0.100 & 0.009  & \textbf{1.306±0.091} & \textbf{0.009 } & 1.679±0.047 & 0.013  & \underline{1.309±0.992} &\underline{ 0.009}  & 1.597±0.098 & 0.010  & 1.323±0.110 & 0.010  & 1.392±0.068 & 0.009  \\
				&       & 100\% & 1.340±0.046 & 0.008  & 19.491±1.953 & 0.046  & 1.307±0.134 & 0.009  & 1.309±0.080 & 0.009  & 1.307±0.046 & 0.010  & 1.443±0.139 & 0.010  & \underline{1.306±0.115} & \underline{0.009}  & 2.671±0.128 & 0.011  & \textbf{1.286±0.041} & \textbf{0.010 } & 1.407±0.082 & 0.009  \\
				\cmidrule{2-23}          & \multirow{3}[2]{*}{\begin{sideways}UniSydney\end{sideways}} & 10\%  & \textbf{0.603±0.026} & \textbf{0.005 } & 32.448±1.203 & 0.115  & \underline{0.783±0.067} & \underline{0.005}  & 88.625±3.856 & 0.912  & 79.612±4.836 & 0.887  & 9.686±0.518 & 0.073  & 1.043±0.111 & 0.008  & 4.669±0.324 & 0.045  & 85.208±6.658 & 0.877  & 60.993±5.710 & 0.627  \\
				&       & 50\%  & \textbf{0.549±0.046} & \textbf{0.004 } & 32.444±1.123 & 0.114  & \underline{0.575±0.018} & \underline{0.004}  & 0.647±0.066 & 0.005  & 1.405±0.113 & 0.010  & 1.376±0.117 & 0.011  & 0.653±0.061 & 0.005  & 15.331±1.233 & 0.152  & 0.586±0.053 & 0.005  & 0.644±0.033 & 0.005  \\
				&       & 100\% & \textbf{0.440±0.042} & \textbf{0.003 } & 32.442±1.296 & 0.114  & 0.513±0.019 & 0.004  & 0.642±0.059 & 0.005  & \underline{0.492±0.038} & \underline{0.004} & 0.809±0.045 & 0.007  & 0.634±0.023 & 0.005  & 9.646±0.522 & 0.099  & 0.613±0.018 & 0.005  & 0.564±0.044 & 0.004  \\
				\midrule
				\multirow{9}[6]{*}{HR (bpm)} & \multirow{3}[2]{*}{\begin{sideways}BIDMC\end{sideways}} & 10\%  & 3.241±0.174 & 0.022  & 4.988±0.342 & 0.035  & 3.857±0.250 & 0.022  & 12.392±0.548 & 0.106  & 3.461±0.103 & 0.022  & 3.215±0.195 & 0.024  & 13.538±1.214 & 0.115  & 17.537±1.171 & 0.145  & \textbf{2.795±0.222} & \textbf{0.019 } & \underline{3.194±0.156} & \underline{0.024}  \\
				&       & 50\%  & \textbf{2.161±0.127} & \textbf{0.015 } & 3.162±0.252 & 0.021  & 3.340±0.168 & 0.018  & 11.453±0.335 & 0.099  & 2.764±0.249 & 0.017  & 2.375±0.255 & 0.018  & 13.535±0.560 & 0.115  & 5.066±0.355 & 0.037  & \underline{2.161±0.134} & \underline{0.015}  & 2.357±0.183 & 0.017  \\
				&       & 100\% & \textbf{1.657±0.131} & \textbf{0.012 } & 2.694±0.139 & 0.019  & 3.057±0.214 & 0.017  & 11.093±0.279 & 0.081  & 1.973±0.210 & 0.014  & 2.120±0.162 & 0.017  & 13.535±1.148 & 0.115  & 5.832±0.602 & 0.044  & \underline{1.911±0.048} & \underline{0.014}  & 2.031±0.170 & 0.014  \\
				\cmidrule{2-23}          & \multirow{3}[2]{*}{\begin{sideways}IEEEPPG\end{sideways}} & 10\%  & \underline{18.370±1.445} & \underline{0.099}  & 34.884±1.778 & 0.262  & 29.328±1.773 & 0.230  & 29.461±2.674 & 0.233  & 52.246±3.552 & 0.348  & 23.141±2.084 & 0.165  & 29.103±2.484 & 0.230  & 30.764±1.097 & 0.234  & \textbf{18.363±1.494} & \textbf{0.153 } & 24.006±2.582 & 0.181  \\
				&       & 50\%  & \textbf{11.492±1.156} & \textbf{0.062 } & 28.699±2.332 & 0.214  & 29.338±2.339 & 0.216  & 27.672±0.732 & 0.216  & 18.957±2.162 & 0.122  & 16.551±1.140 & 0.100  & 28.989±2.405 & 0.228  & 36.151±3.384 & 0.261  & \underline{13.362±1.223} & \underline{0.079}  & 18.498±1.831 & 0.118  \\
				&       & 100\% & \underline{13.831±0.846} &\underline{ 0.079}  & 27.063±1.224 & 0.209  & 27.672±2.873 & 0.231  & 27.750±0.708 & 0.216  & 16.345±0.715 & 0.101  & 14.006±0.481 & 0.087  & 28.976±2.926 & 0.228  & 29.233±1.268 & 0.229  & \textbf{9.959±0.718} & \textbf{0.057 } & 16.980±1.597 & 0.106  \\
				\cmidrule{2-23}          & \multirow{3}[2]{*}{\begin{sideways}UniSydney\end{sideways}} & 10\%  & \underline{6.543±0.509} & \underline{0.067}  & 11.940±0.769 & 0.108  & 11.629±0.334 & 0.106  & 11.610±0.381 & 0.107  & 7.949±0.561 & 0.087  & \textbf{6.275±0.207} & \textbf{0.065 } & 11.632±2.703 & 0.106  & 20.651±1.529 & 0.214  & 7.230±0.653 & 0.076  & 7.341±0.690 & 0.080  \\
				&       & 50\%  & \textbf{4.039±0.143} & \textbf{0.039 } & 10.918±0.399 & 0.107  & 11.629±1.151 & 0.106  & 11.619±1.213 & 0.107  & 6.931±0.570 & 0.072  & \underline{5.326±0.486} &\underline{ 0.054}  & 11.630±1.092 & 0.106  & 11.783±0.684 & 0.108  & 5.793±0.625 & 0.059  & 6.416±0.623 & 0.067  \\
				&       & 100\% & \textbf{3.601±0.296} & \textbf{0.034 } & 11.620±0.951 & 0.106  & 11.629±0.475 & 0.106  & 11.617±1.128 & 0.107  & 6.419±0.351 & 0.065  & 5.057±0.329 & 0.051  & 11.629±0.552 & 0.106  & 11.621±0.821 & 0.106  & \underline{4.867±0.151} & \underline{0.049 } & 5.372±0.253 & 0.054  \\
				\midrule
				\multirow{9}[6]{*}{RR (bpm)} & \multirow{3}[2]{*}{\begin{sideways}BIDMC\end{sideways}} & 10\%  & \textbf{1.484±0.047} & \textbf{0.075 } & 2.718±0.073 & 0.159  & 2.609±0.146 & 0.142  & 3.024±0.256 & 0.183  & 2.123±0.183 & 0.114  & 1.608±0.058 & 0.084  & 3.098±0.184 & 0.193  & 3.273±0.252 & 0.196  & 1.621±0.102 & 0.087  & \underline{1.558±0.087} & \underline{0.084}  \\
				&       & 50\%  & \textbf{1.222±0.096} & \textbf{0.065 } & 1.534±0.116 & 0.076  & 2.037±0.117 & 0.101  & 2.970±0.134 & 0.178  & 1.622±0.075 & 0.083  & 1.310±0.107 & 0.070  & 2.934±0.240 & 0.175  & 3.136±0.238 & 0.187  & 1.302±0.056 & 0.069  & \underline{1.283±0.111} &\underline{ 0.062}  \\
				&       & 100\% & \underline{1.016±0.062} & \underline{0.044}  & 1.410±0.137 & 0.071  & 1.853±0.163 & 0.094  & 2.333±0.193 & 0.114  & 1.440±0.130 & 0.077  & 1.197±0.118 & 0.066  & 3.101±0.158 & 0.189  & 2.967±0.099 & 0.176  & 1.197±0.109 & 0.065  & \textbf{0.935±0.065} & \textbf{0.054 } \\
				\cmidrule{2-23}          & \multirow{3}[2]{*}{\begin{sideways}CapnoBase\end{sideways}} & 10\%  & \textbf{4.657±0.447} & \textbf{0.206 } & 7.937±0.793 & 0.364  & \underline{7.790±0.652} & \underline{0.395}  & 20.185±1.283 & 1.090  & 18.702±1.174 & 0.954  & 18.873±1.490 & 1.114  & 11.247±1.107 & 0.359  & 9.016±0.523 & 0.379  & 19.921±0.900 & 0.967  & 17.823±0.630 & 0.836  \\
				&       & 50\%  & \textbf{4.055±0.254} & \textbf{0.190 } & 7.139±0.290 & 0.310  & 6.056±0.158 & 0.271  & 18.116±1.123 & 1.001  & 13.132±1.068 & 0.696  & \underline{4.125±0.128} & \underline{0.205}  & 7.492±0.709 & 0.376  & 8.671±0.763 & 0.325  & 15.729±0.517 & 0.784  & 7.782±0.199 & 0.366  \\
				&       & 100\% & \textbf{3.440±0.146} & \textbf{0.158 } & 5.361±0.431 & 0.237  & 5.383±0.492 & 0.232  & 14.057±1.223 & 0.731  & 5.375±0.205 & 0.241  & \underline{3.494±0.152} & \underline{0.168}  & 6.591±0.563 & 0.321  & 6.837±0.528 & 0.272  & 9.215±0.616 & 0.451  & 3.793±0.369 & 0.149  \\
				\cmidrule{2-23}          & \multirow{3}[2]{*}{\begin{sideways}PWDB\end{sideways}} & 10\%  & \underline{11.548±0.782} & \underline{0.486}  & 16.629±0.442 & 0.909  & 16.222±0.659 & 0.868  & 18.615±1.124 & 0.629  & 14.279±0.513 & 0.662  & \textbf{9.080±0.328} & \textbf{0.352 } & 16.226±1.667 & 0.864  & 23.218±2.792 & 0.656  & 14.899±4.795 & 0.539  & 14.595±1.330 & 0.762  \\
				&       & 50\%  & \textbf{5.060±0.183} & \textbf{0.203 } & 16.282±0.924 & 0.837  & 16.217±0.885 & 0.873  & 16.173±1.551 & 0.894  & 8.446±0.271 & 0.363  & \underline{5.300±0.507} & \underline{0.252}  & 16.215±1.403 & 0.878  & 17.518±1.611 & 0.757  & 6.207±0.347 & 0.229  & 11.102±0.657 & 0.535  \\
				&       & 100\% & 4.878±0.328 & 0.157  & 16.017±1.095 & 0.826  & 16.214±1.278 & 0.882  & 16.057±0.531 & 0.871  & 7.688±0.649 & 0.348  & \textbf{4.040±0.362} & \textbf{0.158 } & 16.216±0.518 & 0.876  & 16.761±0.460 & 1.018  & \underline{4.584±0.252} & \underline{0.193}  & 6.493±0.525 & 0.330  \\
				\midrule
				\multirow{9}[6]{*}{SBP (mmHg)} & \multirow{3}[2]{*}{\begin{sideways}NBPE\end{sideways}} & 10\%  & \textbf{8.204±0.417} & \textbf{0.049 } & 17.560±1.572 & 0.106  & 15.526±0.389 & 0.093  & 14.762±1.582 & 0.090  & 14.035±0.774 & 0.080  & 9.770±0.765 & 0.056  & 15.559±1.503 & 0.094  & 19.740±0.818 & 0.122  & \underline{9.239±0.910} & \underline{0.051}  & 11.113±0.322 & 0.065  \\
				&       & 50\%  & \textbf{5.235±0.511} & \textbf{0.030 } & 14.871±0.566 & 0.091  & 15.499±0.879 & 0.093  & 14.497±1.419 & 0.086  & 12.445±1.201 & 0.071  & 6.768±0.365 & 0.038  & 15.496±5.671 & 0.093  & 17.706±1.918 & 0.103  & \underline{6.120±0.541} & \underline{0.034}  & 6.196±0.234 & 0.032  \\
				&       & 100\% & \textbf{4.184±0.182} & \textbf{0.018 } & 14.270±1.512 & 0.086  & 15.492±1.641 & 0.093  & 14.426±1.312 & 0.087  & 10.749±0.529 & 0.060  & 6.373±0.556 & 0.034  & 15.493±1.121 & 0.093  & 16.071±0.839 & 0.096  & 5.203±0.206 & 0.028  & \underline{4.716±0.148} & \underline{0.026}  \\
				\cmidrule{2-23}          & \multirow{3}[2]{*}{\begin{sideways}PPGBP\end{sideways}} & 10\%  & \textbf{20.987±2.227} & \textbf{0.142 } & \underline{21.159±1.552} & \underline{0.143}  & 21.492±1.376 & 0.145  & 54.353±5.512 & 0.424  & 33.275±0.836 & 0.209  & 23.517±1.517 & 0.156  & 43.899±1.828 & 0.284  & 29.455±1.282 & 0.187  & 23.290±1.842 & 0.145  & 22.087±2.108 & 0.139  \\
				&       & 50\%  & \textbf{17.451±0.520} & \textbf{0.105 } & 21.422±1.128 & 0.142  & 22.812±0.677 & 0.145  & 21.039±1.508 & 0.136  & 21.039±1.756 & 0.137  & \underline{17.800±1.629} & \underline{0.111}  & 20.908±1.265 & 0.134  & 33.481±1.340 & 0.205  & 18.737±1.881 & 0.120  & 19.296±4.982 & 0.122  \\
				&       & 100\% & \textbf{15.858±1.580} & \textbf{0.098 } & 21.234±1.881 & 0.139  & 21.026±1.017 & 0.134  & 20.676±2.190 & 0.131  & 20.524±1.392 & 0.131  & \underline{17.225±0.603} &\underline{ 0.109}  & 21.019±1.885 & 0.135  & 42.606±4.086 & 0.259  & 17.824±0.996 & 0.114  & 17.760±0.814 & 0.111  \\
				\cmidrule{2-23}          & \multirow{3}[2]{*}{\begin{sideways}UniSydney\end{sideways}} & 10\%  & \textbf{8.334±0.286} & 0.052  & 13.074±1.242 & 0.080  & 13.044±0.568 & 0.079  & 13.020±1.041 & 0.078  & 12.241±1.307 & 0.076  & \underline{8.386±0.545} & \underline{0.051 } & 13.052±0.386 & 0.080  & 13.362±1.145 & 0.083  & 10.096±0.719 & 0.063  & 9.962±0.956 & 0.061  \\
				&       & 50\%  & \textbf{5.271±0.311} & \textbf{0.030 } & 13.036±0.839 & 0.079  & 13.056±0.637 & 0.080  & 13.006±5.326 & 0.078  & 10.241±0.475 & 0.064  & 7.266±0.204 & 0.045  & 13.042±1.161 & 0.079  & 13.049±0.461 & 0.080  & \underline{7.132±0.614} & \underline{0.043}  & 7.918±0.735 & 0.049  \\
				&       & 100\% & \textbf{4.250±0.342} & \textbf{0.023 } & 13.043±0.898 & 0.079  & 13.042±0.945 & 0.079  & 13.003±1.357 & 0.078  & 9.665±0.848 & 0.062  & 7.140±0.544 & 0.044  & 13.042±1.299 & 0.079  & 13.701±0.935 & 0.088  & \underline{6.349±0.443} & \underline{0.038}  & 6.361±0.678 & 0.039  \\
				\midrule
				\multirow{9}[6]{*}{DBP (mmHg)} & \multirow{3}[2]{*}{\begin{sideways}NBPE\end{sideways}} & 10\%  & \textbf{4.754±0.197} & \textbf{0.053 } & 7.741±0.413 & 0.089  & 7.332±0.199 & 0.085  & 7.017±0.661 & 0.078  & 7.566±0.525 & 0.080  & 6.692±0.347 & 0.072  & 7.329±0.663 & 0.085  & 9.177±0.460 & 0.100  & \underline{5.390±0.446} & \underline{0.056}  & 5.819±0.342 & 0.062  \\
				&       & 50\%  & \textbf{3.508±0.121} & \textbf{0.035 } & 7.667±0.537 & 0.088  & 7.331±0.299 & 0.085  & 6.985±0.214 & 0.079  & 5.807±0.206 & 0.064  & 3.759±0.361 & 0.039  & 7.335±0.458 & 0.085  & 7.549±0.558 & 0.086  & \underline{3.550±0.226} & \underline{0.035}  & 4.054±0.213 & 0.043  \\
				&       & 100\% & \underline{2.916±0.188} & \underline{0.027}  & 7.135±0.598 & 0.081  & 7.329±0.227 & 0.085  & 7.074±0.393 & 0.078  & 4.888±0.369 & 0.053  & 3.260±0.088 & 0.034  & 7.329±0.542 & 0.085  & 7.335±0.230 & 0.086  & \textbf{2.914±0.140} & \textbf{0.028 } & 3.245±0.328 & 0.031  \\
				\cmidrule{2-23}          & \multirow{3}[2]{*}{\begin{sideways}PPGBP\end{sideways}} & 10\%  & 14.467±1.566 & 0.158  & 23.950±2.140 & 0.281  & \textbf{12.110±0.869} & \textbf{0.132 } & 34.100±3.273 & 0.473  & 17.805±0.947 & 0.183  & \underline{12.329±0.996} & \underline{0.139}  & 12.344±1.246 & 0.127  & 33.292±2.031 & 0.355  & 37.301±3.878 & 0.473  & 13.934±0.434 & 0.151  \\
				&       & 50\%  & \underline{11.445±0.443} &\underline{ 0.120}  & 13.391±1.139 & 0.149  & 12.148±1.064 & 0.130  & 11.929±1.154 & 0.132  & 13.050±0.744 & 0.147  & 11.645±0.650 & 0.128  & 12.138±0.446 & 0.131  & 14.024±1.497 & 0.179  & 11.720±0.762 & 0.127  & \textbf{11.140±0.928} & \textbf{0.122 } \\
				&       & 100\% & \textbf{10.659±0.507} & \textbf{0.115 } & 12.963±1.152 & 0.142  & 12.142±1.051 & 0.131  & 11.967±0.911 & 0.133  & 11.893±0.453 & 0.128  & 10.781±1.060 & 0.116  & 12.149±1.217 & 0.130  & 13.324±0.582 & 0.138  & 10.903±1.016 & 0.114  & \underline{10.726±0.394} & \underline{0.118}  \\
				\cmidrule{2-23}          & \multirow{3}[2]{*}{\begin{sideways}UniSydney\end{sideways}} & 10\%  & \textbf{7.863±0.558} & \textbf{0.087 } & 58.312±2.179 & 0.757  & 57.024±2.355 & 0.743  & 75.731±5.445 & 0.997  & 74.241±4.096 & 0.977  & 75.127±4.242 & 0.998  & 60.359±2.800 & 0.788  & \underline{15.003±0.913} & \underline{0.171}  & 74.821±7.796 & 0.985  & 71.890±7.581 & 0.945  \\
				&       & 50\%  & \textbf{5.494±0.301} & \textbf{0.065 } & \underline{9.939±0.307} & \underline{0.114}  & 30.069±1.290 & 0.374  & 75.473±2.173 & 0.993  & 71.256±7.555 & 0.936  & 54.491±4.954 & 0.708  & 23.869±1.846 & 0.285  & 20.421±1.347 & 0.241  & 72.890±2.383 & 0.959  & 66.738±7.166 & 0.877  \\
				&       & 100\% & \textbf{4.361±0.171} & \textbf{0.051 } & 10.473±1.049 & 0.121  & 8.445±0.344 & 0.093  & 74.387±2.079 & 0.979  & 65.818±5.341 & 0.863  & 14.879±1.323 & 0.156  & \underline{8.301±0.457} & \underline{0.090}  & 16.979±0.945 & 0.197  & 69.330±5.783 & 0.911  & 56.018±1.691 & 0.734  \\
				\bottomrule
			\end{tabular}%
		}
		\label{tab4}%
	\end{table*}%
	
	Experiments were conducted with different amounts of training data in downstream tasks. \Cref{tab4} shows that the proposed method significantly outperforms most currently existing methods with limited data. Compared to the best performance of all participating methods, the proposed method achieved an average improvement of 5.70\% in RMSE when only 10\% of the data was used for training. This indicates that the initialization weights learned through the pre-training process perform well with limited data. It is worth noting that there was also an improvement in databases with limited data, such as IEEEPPG and PPGBP, further demonstrating the effectiveness of the pre-training process. After injecting external continuous features and inflating the initial feature dimensions, as well as benefiting from the good global semantic feature extraction and local context representation obtained by simultaneously modeling temporal dependencies and encoding reconstructions at a coarse-grained fragment level, the proposed method achieved an average improvement of 7.95\% compared to the best results on the whole training set (100\%).
	
	Furthermore, we found that the proposed method performed best with 50\% of the training data, achieving an average improvement of 11.46\% in RMSE. This is due to the benefits of dual-process encoder transfer, which provides a general data distribution for observation in downstream tasks, maximizing the effectiveness of the DPT method. However, there is still a gap in the results when using all the training data. While some generative models like TSSequencer perform well in the case of scarce PPG training samples (10\%), they are challenging to train further with more observational data, either converging too early or too slowly. This issue does not exist in convolutional neural networks such as OmniScale, which is one of the reasons we chose neural networks as the backbone.
	
	\subsection{Exploration of CTFGA's best $\gamma$ and $T$}
	\label{Exploration of CTFGA's best γ and T}
	
\begin{figure}
		\centering
		\includegraphics[width=60mm]{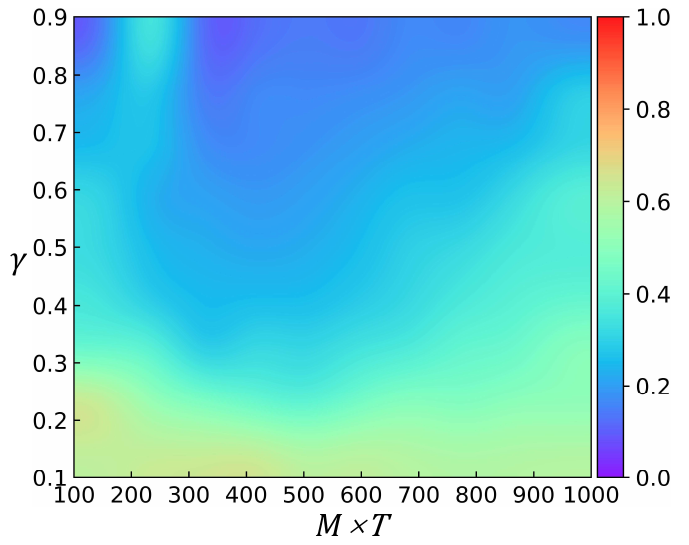}
		\caption{Pretraining Validation Loss Results at Different $\gamma$ and $T$.}
		\label{fig10}
	\end{figure}

	In TS2TC, the most critical issue is how to choose the length of the PPG sequence as an instance and how to partition the 'past-anchor-future' segments when generating anchor sequences across temporal, i.e., how to determine the instance sampling duration $T$ and the partition ratio $\gamma$. The conventional intuition is that as $\gamma$ and $T$ increase, the difficulty of the pre-training task also increases, typically manifested as an increase in reconstruction loss. However, through experiments, we found that this relationship is not simply linear. As shown in \cref{fig10}, there is a decrease in task difficulty on moderately sized instance lengths $M\times T$, indicating that discriminative features may be learned at appropriate values of $\gamma$ and $T$. Therefore, this unconventional phenomenon guides us to experiment further, verifying the impact of $\gamma$ and $T$ on downstream tasks and selecting the optimal parameters, rather than simply increasing the instance length and pretext task difficulty.
	
\begin{figure*}[h]
		\centering
		\includegraphics[width=130mm]{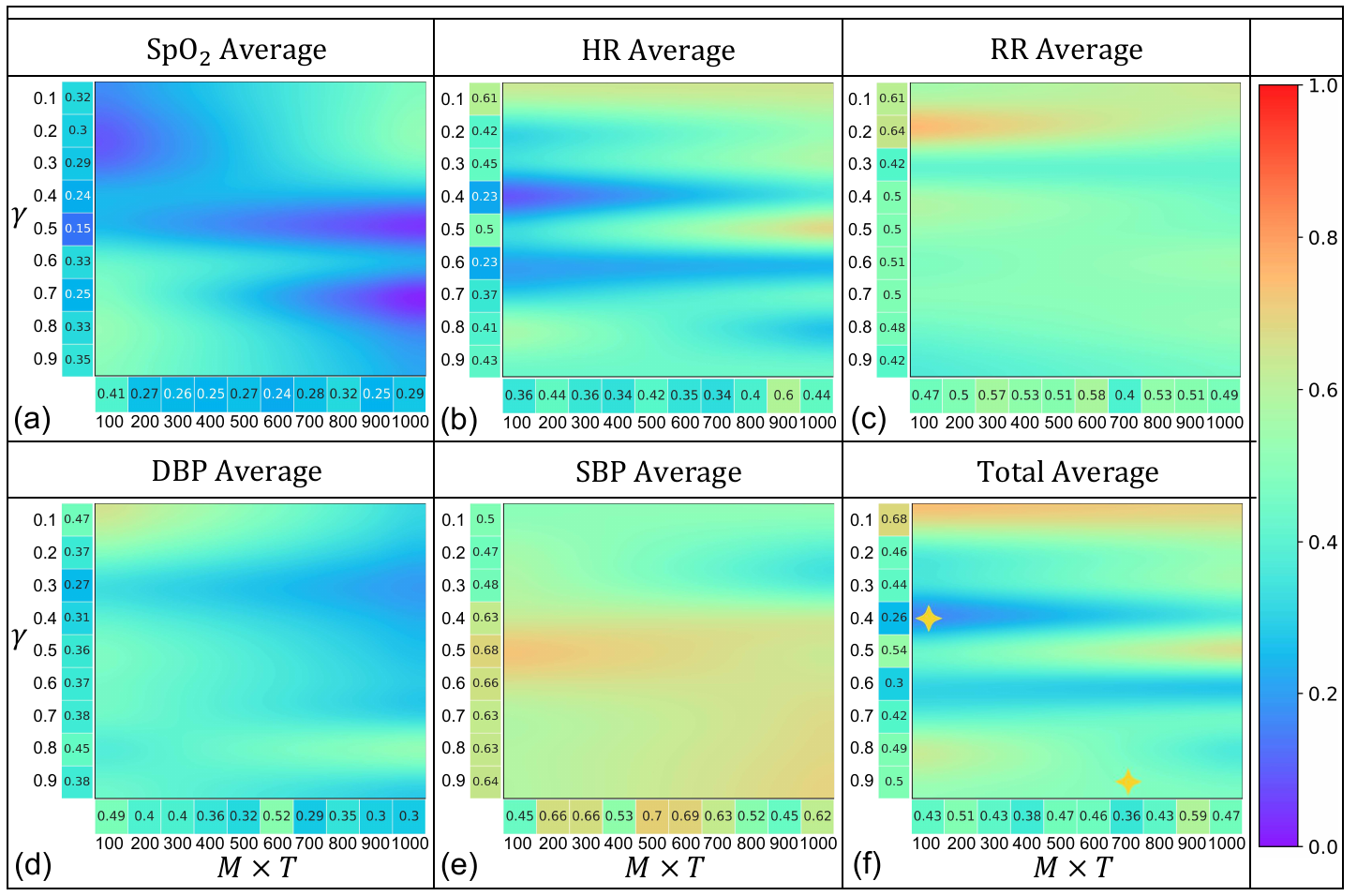}
		\caption{Performance of different $\gamma$ and $T$ values in downstream tasks, with asterisks indicating the best performance.}
		\label{fig11}
	\end{figure*}

	As shown in \cref{fig11}, to facilitate comparisons across different tasks, we used the Min-Max Normalization method to scale the training results to $\left [ 0,1 \right ]$. \cref{fig11}(a)-(e) show the averages of the results for the three datasets mentioned in \cref{tab1} for each prediction task, with the horizontal and vertical bars indicating the mean values across rows and columns, respectively, providing a larger-scale view of the significant impact of $\gamma$ and $T$ on downstream tasks. To use fixed and appropriate parameters in a general framework, we averaged all the results to obtain \cref{fig11}(f), which shows that the overall best performance is achieved when $\gamma$ is set to 0.4, and $T$ is set to 5.6, corresponding to a complete instance length of $M\times T=700$ and an anchor sequence segment of 280.
	
	\subsection{Ablation study: VMD, MSDI, and ZDL}
	\label{Ablation study: VMD, MSDI, and ZDL}
	
	To explore the effectiveness of variational mode decomposition (VMD), multi-stage derivative staged injection (MSDI), and zero decoder layer (ZDL), we conducted ablation experiments on the whole training set (\cref{tab5}), all with self-supervised pre-training weights. OCTFGA is the original method without dilation and injection, and it can be seen that both VMD and MSDI have improved performance in all tasks. More notably, ZDL significantly improves results, mainly because we did not screen the original data based on signal quality or perform bandpass filtering or denoising operations on the signals. This indirectly confirms that observing harmful data distributions can lead to biases in the results, while ZDL can preserve beneficial information during feature compression and decoding, thereby increasing the robustness of the decoding process.
	
	\begin{table}[h]
		\centering
		\caption{Ablation of VMD, MSDI, and ZDL.}
		\resizebox{\linewidth}{!}{
			\begin{tabular}{cp{6em}p{6em}cp{6em}cp{6em}cp{6em}c}
				\toprule\toprule
				& Model & \multicolumn{2}{c}{OCTFG} & \multicolumn{2}{c}{OCTFG+VMD} & \multicolumn{2}{c}{OCTFG+VMD+MSDI} & \multicolumn{2}{c}{OCTFG+VMD+MSDI+ZDL} \\
				\midrule
				& Database & \multicolumn{1}{c}{MAE}   & \multicolumn{1}{c}{MAPE} & \multicolumn{1}{c}{MAE}   & \multicolumn{1}{c}{MAPE} & \multicolumn{1}{c}{MAE}   & \multicolumn{1}{c}{MAPE} & \multicolumn{1}{c}{MAE}   & \multicolumn{1}{c}{MAPE} \\
				\midrule
				\multicolumn{1}{c}{\multirow{3}[2]{*}{$\text{SpO}_{2}$}} & BIDMC & 1.698±0.252 & 0.018 & 1.695±0.111 & 0.018 & 1.614±0.047 & 0.016 & 1.007±0.049 & 0.010 \\
				& UniGreifswald & 1.170±0.030 & 0.012 & 1.144±0.099 & 0.012 & 1.085±0.069 & 0.012 & 0.947±0.022 & 0.010 \\
				& UniSydney & 8.201±1.602 & 0.084 & 5.929±0.407 & 0.061 & 4.931±0.348 & 0.056 & 0.467±0.018 & 0.005 \\
				\midrule
				\multicolumn{1}{c}{\multirow{3}[2]{*}{HR}} & BIDMC & 2.345±0.040 & 0.028 & 2.205±0.039 & 0.026 & 2.109±0.189 & 0.026 & 1.879±0.074 & 0.022 \\
				& IEEEPPG & 16.491±0.484 & 0.145 & 16.280±0.111 & 0.146 & 16.245±1.404 & 0.146 & 15.293±0.241 & 0.149 \\
				& UniSydney & 6.002±0.792 & 0.083 & 5.746±0.855 & 0.078 & 5.48±0.525 & 0.076 & 4.930±0.083 & 0.067 \\
				\midrule
				\multicolumn{1}{c}{\multirow{3}[2]{*}{RR}} & BIDMC & 1.325±0.246 & 0.101 & 1.247±0.059 & 0.098 & 1.239±0.049 & 0.093 & 0.961±0.045 & 0.075 \\
				& PWDB  & 9.738±1.938 & 0.472 & 9.608±0.123 & 0.476 & 9.403±0.479 & 0.478 & 7.925±0.318 & 0.486 \\
				& CapnoBase & 3.744±0.438 & 0.256 & 3.732±0.348 & 0.241 & 3.683±0.093 & 0.231 & 3.258±0.108 & 0.206 \\
				\midrule
				\multicolumn{1}{c}{\multirow{3}[2]{*}{DBP}} & UniSydney & 71.314±6.174 & 0.945 & 55.938±2.133 & 0.663 & 42.869±2.732 & 0.487 & 6.210±0.143 & 0.087 \\
				& PPGBP & 12.908±0.416 & 0.181 & 12.490±1.719 & 0.174 & 12.421±0.694 & 0.171 & 11.280±0.074 & 0.158 \\
				& NBPE  & 5.046±0.390 & 0.072 & 4.720±0.566 & 0.068 & 4.412±0.204 & 0.063 & 3.757±0.048 & 0.053 \\
				\midrule
				\multicolumn{1}{c}{\multirow{3}[2]{*}{SBP}} & UniSydney & 6.689±0.392 & 0.057 & 6.463±0.483 & 0.057 & 6.334±0.536 & 0.055 & 6.006±0.278 & 0.052 \\
				& PPGBP & 17.821±2.120 & 0.144 & 17.457±1.829 & 0.144 & 17.309±1.633 & 0.144 & 16.639±0.737 & 0.142 \\
				& NBPE  & 11.616±0.392 & 0.085 & 11.041±0.083 & 0.073 & 9.675±0.643 & 0.068 & 6.065±0.170 & 0.049 \\
				\bottomrule
			\end{tabular}%
		}
		\label{tab5}%
	\end{table}%
	
	\subsection{Explanation of the contribution}
	\label{Explanation of the contribution}
	
	As described in \cref{Proposed TS2TC framework}, R-squared is used to assess the fit of a regression model, indicating the extent to which the independent variables $\hat{y}_{j}$ explain the dependent variable $g({\hat{y}})$. The independent variables come from the predicted outputs of three domains, while the dependent variable is the overall output of TS2TC. Therefore, R-squared can be used to observe the effectiveness of the fusion method. \Cref{tab6} shows that after the components of the TS2TC framework collaborate, they exhibit a high degree of explanation for the reference labels representing ground truth. However, there is a slight decrease in the BP task, indicating that there is still potential information in PPG waiting to be further explored.
	\begin{table}[h]
		\centering
		\caption{Parameters of the TS2TC framework based on OLS fusion across different tasks.}
		\resizebox{\linewidth}{!}{
			\begin{threeparttable}
				\begin{tabular}{cccccc}
					\toprule\toprule
					& Database & R-squared & $\phi_{1}$    & $\phi_{2}$    & $\phi_{3}$ \\
					\midrule
					\multirow{3}[2]{*}{$\text{SpO}_{2}$} & BIDMC & 0.916±0.034 & 0.755±0.014 & -0.106±0.012 & 0.755±0.099 \\
					& UniGreifswald & 0.906±0.027 & 0.853±0.012 & -0.190±0.016 & 0.853±0.063 \\
					& UniSydney & 0.918±0.038 & 0.848±0.026 & 0.170±0.028 & 0.848±0.081 \\
					\midrule
					\multirow{3}[2]{*}{HR} & BIDMC & 0.953±0.036 & 0.997±0.031 & 0.129±0.066 & 0.997±0.072 \\
					& IEEEPPG & 0.969±0.022 & 0.943±0.033 & -0.115±0.062 & 0.943±0.126 \\
					& UniSydney & 0.934±0.029 & 0.932±0.014 & -0.296±0.036 & 0.932±0.090 \\
					\midrule
					\multirow{3}[2]{*}{RR} & BIDMC & 0.879±0.031 & 1.031±0.014 & 0.022±0.032 & 1.031±0.027 \\
					& PWDB  & 0.972±0.018 & 0.817±0.029 & 0.022±0.064 & 0.817±0.012 \\
					& CapnoBase & 0.985±0.013 & 0.989±0.010 & 0.086±0.060 & 0.989±0.135 \\
					\midrule
					\multirow{3}[2]{*}{DBP} & UniSydney & 0.607±0.024 & 0.846±0.026 & 0.403±0.080 & 0.846±0.096 \\
					& PPGBP & 0.953±0.025 & 0.917±0.011 & 0.208±0.046 & 0.917±0.036 \\
					& NBPE  & 0.879±0.033 & 0.835±0.015 & 0.207±0.032 & 0.835±0.102 \\
					\midrule
					\multirow{3}[2]{*}{SBP} & UniSydney & 0.820±0.016 & 0.971±0.037 & 0.05±0.044 & 0.971±0.048 \\
					& PPGBP & 0.944±0.027 & 0.911±0.016 & 0.006±0.118 & 0.911±0.117 \\
					& NBPE  & 0.869±0.010 & 0.817±0.033 & 0.278±0.020 & 0.817±0.013 \\
					\bottomrule
				\end{tabular}%
				\begin{tablenotes}
					\footnotesize
					\item[1] $\phi_{j}$ represents the interaction contribution of different domains in the TS2TC framework, where $\phi_{1}$ denotes the TS2TC temporal domain, $\phi_{2}$ denotes the TS2TC spectrogram domain, $\phi_{3}$ denotes the temporal-spectrogram mixed domain.
					\item[2] R-squared indicates the explanatory extend of the coordinated domains for the reference labels.
				\end{tablenotes}
			\end{threeparttable}
		}
		\label{tab6}%
	\end{table}%
	
	Additionally, by observing $\phi_{j}$ of TS2TC, we can see the relative contributions of different components to the results. The larger the absolute value of the weight, the higher the impact of the predicted value of that component on the results. This also demonstrates the significant contribution of the original temporal domain to various physiological parameter tasks. Moreover, in some tasks, such as DBP prediction, the spectrogram domain also plays an important role, inspiring new ideas for blood pressure prediction.
	
\begin{figure}
		\centering
		\includegraphics[width=80mm]{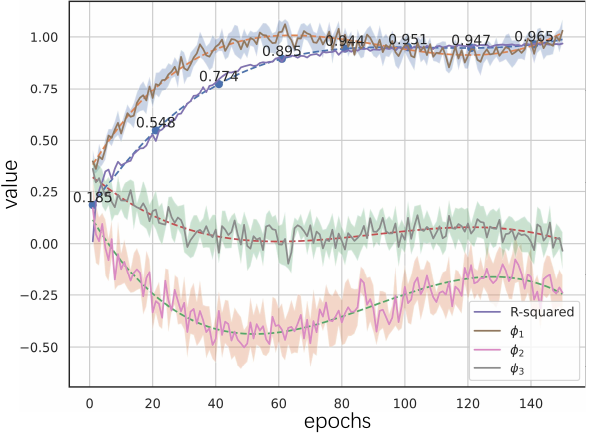}
		\caption{Analysis of the contributions of components in the TS2TC framework on IEEEPPG.}
		\label{fig12}
	\end{figure}

	\cref{fig12} shows the evolution of R-squared and $\phi_{j}$ for the components of the TS2TC framework during the training process in the experiment where 10\% of the IEEEPPG data was used. It can be observed that, as the neural network learns, the spectrogram domain's contribution initially dominates, followed by the temporal domain taking the lead, further highlighting the contribution of the original temporal domain. Additionally, the explanatory power of each component as independent variables for the standard results as dependent variables steadily increases, indicating that the network learns discriminative features.
	
	\subsection{Ternary fusion: linear or nonlinear}
	\label{Ternary Fusion: Linear or Nonlinear}
	In the ternary fusion strategy, we used a simple ordinary least squares method to determine $\phi_{j}$. Generally, ordinary least squares method is not effective for fitting nonlinear relationships, sensitive to outliers, and not suitable for complex data patterns, but it is suitable for modeling and predicting simple linear relationships.
	
	Consider the local situation under the ordinary least squares method in \cref{eq25}: $g(\hat{y})=\hat{y}_1$. That is, without fusion, only the temporal domain predictions of TS2TC are used. In this case, $\phi_{0} = \phi_{2} = \phi_{3} = 0$, and $\phi_{1} = 1$. However, as shown in \cref{tab2}, the temporal domain does not perform well in all tasks, and there is no prior knowledge to guarantee the selection of the best domain components on new samples to achieve good performance. Therefore, it is necessary to weigh the advantages of different domains using weights, which is the reason for using ordinary least squares method, aiming to provide a simple and well-explained form of collaboration and fusion. Furthermore, based on the balancing idea in \cref{eq25}, more nonlinear dynamic weight allocation methods can be explored in the future to further improve the performance of the TS2TC framework.
	
	Since we use ordinary least squares method in a system of three complex nonlinear modeling systems, namely neural networks, the function is only to weigh the advantages of complex nonlinear systems, as the predictions of each system are refitted in a linear form according to their contributions. The purpose is not to achieve the best result for the fusion system, but rather to balance the advantages of complex nonlinear systems. This is also the reason why ordinary least squares method is effective.
	
	\subsection{Estimating latent physiological parameters: performance evaluation}
	\label{Estimating Latent Physiological Parameters: Performance Evaluation}
	
	Conventional methods for blood glucose monitoring involve fingerstick measurements and costly continuous glucose monitoring (CGM) devices. However, fingerstick measurements are known to be painful, uncomfortable, and expensive. Moreover, this approach lacks the capability to continuously monitor blood glucose levels. CGM devices require inserting a probe into the subcutaneous tissue for assessing blood glucose levels, necessitating frequent replacements and causing discomfort and pain. Hence, there is a pressing need for the development of an affordable, non-invasive method for blood glucose monitoring in diabetes management.
	
	Blood glucose (BG) is among the potential physiological parameters linked to photoplethysmographic (PPG) signals, and the accurate estimation of blood glucose using PPG has emerged as a recent and significant research area \citep{15li2023noninvasive}. However, due to its invasive nature and financial constraints, obtaining a sufficient number of aligned labels poses new challenges for technical methods. We assessed the efficacy of the proposed TS2TC framework with pre-trained weights on the private database, Blood Glucose DataBase (BGDB).
	
	\begin{figure}
		\centering
		\includegraphics[width=50mm]{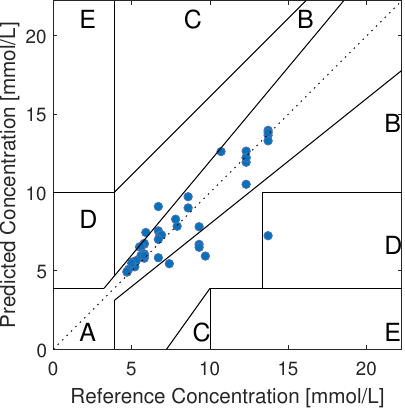}
		\caption{Clarke error grid analysis of the proposed TS2TC.}
		\label{fig13}
	\end{figure}
	
	The BGDB database contains PPG single-channel records from 41 adults with an average age of 53.15 years, each record averaging 5900 data points sampled at 125 Hz, providing blood glucose labels obtained through fingertip blood sampling with mean and standard deviation of 7.02 and 2.86 (mmol/L) respectively. Additionally, 21.95\% of abnormal blood glucose samples with postprandial blood glucose values exceeding 7.8 mmol/L are present. Consistent with the processing described in \cref{Datasets} to \cref{Data preprocessing}, 163 records were obtained after processing for training, with 23 and 47 records used for validation and testing, respectively. For privacy protection, individual identities were removed during experimentation.
	
	\begin{table}[htbp]
		\centering
		\caption{Comparison results of different methods on the BGDB dataset.}
		\resizebox{\linewidth}{!}{
		\begin{tabular}{ccccc}
			\toprule\toprule
			Method & \multicolumn{1}{c}{MAE\newline{} (mmol/L)} & \multicolumn{1}{c}{MAPE\newline{} (\%)} & \multicolumn{1}{c}{Zone A\newline{} (\%)} & \multicolumn{1}{c}{Zone B\newline{} (\%)} \\
			\midrule
			PatchTST & 2.522±0.122 & 35.93  & 42.55  & 53.19  \\
			TSSequencer & 1.312±0.072 & 19.08  & 63.82  & 34.04  \\
			TSPerceiver & 2.261±0.257 & 31.76  & 27.65  & 70.21  \\
			XCM   & 1.509±0.086 & 22.31  & 55.31  & 42.55  \\
			MultiRocket & 1.699±0.102 & 25.24  & 40.42  & 57.44  \\
			TSiTransformer  & 1.387±0.019 & 20.91  & 44.68  & 53.19  \\
			TST   & 1.503±0.105 & 21.44  & 55.31  & 42.55  \\
			OmniScale & 2.749±0.012 & 39.98  & 14.89  & 82.97  \\
			InceptionTime & 1.679±0.132 & 22.57  & 42.55  & 55.31  \\
			TS2TC & 1.273±0.146 & 18.23  & 63.82  & 34.04  \\
			TS2TC* & \textbf{0.922±0.251} & \textbf{11.21 } & \textbf{78.72 } & \textbf{19.14 } \\
			\bottomrule
		\end{tabular}%
	}
		\label{tab7}%
	\end{table}%
	
	To demonstrate the clinical potential of TS2TC, the mean Clarke error grid analysis of ten randomly conducted experiments is shown in \cref{fig13}. The results indicate that 97.86\% of BG prediction samples fall into clinically acceptable zones, demonstrating good consistency between predicted and reference values, thereby instilling confidence in our general framework for estimating physiological parameters.
	
	The results in \cref{tab7} display the predictive performance of different methods on the BGDB dataset. In comparison to these ten methods, our proposed approach exhibits the lowest MAE of 0.922±0.251 mmol/L. Furthermore, in the Clarke error grid analysis, it shows the highest zone A at 78.72\% and the lowest zone B at 19.14\%, indicating TS2TC's potential to explore and estimate underlying physiological parameters associated with PPG in a limited sample set.
	
	\section{Conclusion}
	\label{Conclusion}
	
	This paper introduces a universal generative self-supervised representation learning framework, TS2TC, designed for estimating physiological parameters based on photoplethysmographic (PPG) signals. First, the pretext task of reconstructing the cross-temporal fusion to generate anchor (CTFGA) sequences models both temporal dependency and independent fragment encoding at the coarse-grained level to provide robust global feature extraction capability and robust local semantic context representation. In addition, PPG-derived multi-frequency scale sub-signals and multi-order kinetic derivatives are considered internal features at different semantic levels, respectively, to guide the learning of generalized shared representations. Second, a high-level cognitively-inspired, dual-process transfer (DPT) strategy consisting of a priori dependent autonomous processes and a posteriori observed reasoning processes is designed for taking advantage of shared and specific representations independently and fusionally. Besides encompassing the spectrogram domain, lastly, a novel bilinear temporal-spectrogram fusion method is proposed in the mixed domain, which aligns the implicit representations of different domains and establishes the feature on the multi-source information level fine-grained contextual interactions. Further, a contribution-interpretable ternary fusion strategy is used to complement the temporal-spectrogram multivariate advantage. Evaluation across five tasks related to physiological parameters validates the universality and efficacy of TS2TC. Moving forward, TS2TC will explore additional potential physiological parameters through representation learning pre-training of encoders and will be deployed in practical applications such as home or community health monitoring systems.

	\section*{CRediT authorship contribution statement}
	\textbf{Zexing Zhang:} Conceptualization, Methodology, Software, Writing - original draft. \textbf{Huimin Lu:} Supervision, Funding acquisition, Writing - review \& editing. \textbf{Songzhe Ma:} Writing - review \& editing. \textbf{Jianzhong Peng:}  Software. \textbf{Chenglin Lin:} Validation \& Data curation. \textbf{Niya Li:} Funding acquisition. \textbf{Bingwang Dong:} Visualization.
	
	\section*{Declaration of competing interest}
	The authors declare that they have no known competing financial interests or personal relationships that could have appeared to influence the work reported in this paper.
	
	\section*{Acknowledgments}
	
	This research was supported by the Fundamental Research Funds for the Central Universities JLU (No. 93K172022K15), 2023 Jilin Provincial Development and Reform Commission Industrial Technology Research and Development Project (No. 2023C042-6) and the 2023 Jilin Provincial Department of Education Science and Technology Research Planning Key Project (No. JJKH20230763KJ).
	
	\section*{Data availability}
	Data will be made available on request.

	\bibliographystyle{elsarticle-harv} 
	\bibliography{refer}

\end{document}